\documentclass[aps, reprint, amsmath, amssymb, longbibliography, superscriptaddress]{revtex4-2}

\usepackage[T1]{fontenc}
\usepackage{palatino}
\usepackage{soul}
\usepackage{lipsum}
\soulregister\cite7
\soulregister\ref7
\soulregister\pageref7
\soulregister\ce7
\usepackage{graphicx}
\usepackage[svgnames]{xcolor}
\usepackage{amsmath, bm}
\usepackage[detect-none]{siunitx}
\usepackage{fancyref}
\usepackage[colorlinks,
            linkcolor=DarkGreen,
            urlcolor=Blue,
            citecolor=DarkGreen]{hyperref}
\usepackage[utf8]{inputenc}
\usepackage{textcomp}
\usepackage[version=3]{mhchem}
\usepackage{miller}

\newcommand{\orcidicon}[1]{\href{https://orcid.org/#1}{\includegraphics[height=\fontcharht\font`\B]{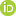}}}

\begin{document}

\title[]{Integration of High-Tc Superconductors with High Q Factor Oxide Mechanical Resonators}

\author{Nicola~\surname{Manca}\,\orcidicon{0000-0002-7768-2500}}
\email{nicola.manca@spin.cnr.it}
\affiliation{CNR-SPIN, C.so F.\,M.~Perrone, 24, 16152 Genova, Italy}
\author{Alexei~\surname{Kalaboukhov}\,\orcidicon{0000-0003-2939-6187}}
\affiliation{Department of Microtechnology and Nanoscience – MC2, Chalmers University of Technology, SE 412 96, Gothenburg, Sweden}
\author{Alejandro~E.~\surname{Plaza}\,\orcidicon{0000-0002-2538-7585}}
\affiliation{CNR-SPIN, C.so F.\,M.~Perrone, 24, 16152 Genova, Italy}
\author{Leon\'elio~\surname{Cichetto}~Jr\,\orcidicon{0000-0002-5894-8852}}
\affiliation{CNR-SPIN, C.so F.\,M.~Perrone, 24, 16152 Genova, Italy}
\author{Emilio~\surname{Bellingeri}\,\orcidicon{0000-0001-7902-0706}}
\affiliation{CNR-SPIN, C.so F.\,M.~Perrone, 24, 16152 Genova, Italy}
\author{Francesco~\surname{Bisio}\,\orcidicon{0000-0003-1776-3023}}
\affiliation{CNR-SPIN, c/o Università di Genova, Via Dodecaneso 33, 16146 Genova, Italy}
\author{Floriana~\surname{Lombardi}\,\orcidicon{0000-0002-3478-3766}}
\affiliation{Department of Microtechnology and Nanoscience – MC2, Chalmers University of Technology, SE 412 96, Gothenburg, Sweden}
\author{Daniele~\surname{Marré}\,\orcidicon{0000-0002-6230-761X}}
\affiliation{Dipartimento di Fisica, Università degli Studi di Genova, 16146 Genova, Italy}
\affiliation{CNR-SPIN, C.so F.\,M.~Perrone, 24, 16152 Genova, Italy}
\author{Luca~\surname{Pellegrino}\,\orcidicon{0000-0003-2051-4837}}
\affiliation{CNR-SPIN, C.so F.\,M.~Perrone, 24, 16152 Genova, Italy}

\begin{abstract}
    Micro-mechanical resonators are building blocks of a variety of applications in basic science and applied electronics. This device technology is mainly based on well-established and reproducible silicon-based fabrication processes with outstanding performances in term of mechanical $Q$ factor and sensitivity to external perturbations. Broadening the functionalities of MEMS by the integration of functional materials is a key step for both applied and fundamental science. However, combining functional materials and silicon-based compounds is challenging. An alternative approach is fabricating MEMS based on complex heterostructures made of materials inherently showing a variety of physical properties such as transition metal oxides. Here, we report on the integration of a high-Tc superconductor \ce{YBa2Cu3O7} (YBCO) with high $Q$ factor micro-bridge resonator made of a single-crystal \ce{LaAlO3} (LAO) thin film. LAO resonators are tensile strained, with a stress of 345\,MPa, show $Q$ factor in the range of tens of thousands, and have low roughness. The topmost YBCO layer deposited by Pulse Laser Deposition shows a superconducting transition starting at 90\,K with zero resistance below 78\,K. This result opens new possibilities towards the development of advanced transducers, such as bolometers or magnetic field detectors, as well as basic science experiments in solid state physics, material science, and quantum opto-mechanics.
\end{abstract}

\maketitle

A distinct feature of many transition metal oxides is their compatible crystal lattice and the consequent opportunity to fabricate epitaxial heterostructures showing emergent properties.\cite{Ramesh2019} Paradigmatic realizations of this capability are conductive interfaces hosting superconducting quasi-2D electron gas,\cite{Zubko2011, Pai2018a, Christensen2019} which can be further manipulated to integrate spin-polarization, ferro-electricity and magnetism.\cite{Stornaiuolo2016, Brehin2023}
Growth of crystalline complex oxides is also possible on pre-patterned templates, such as suspended oxide thin films, where ex-situ etero-epitaxial growth has been demonstrated.\cite{Biasotti2010, Pellegrino2012}. These full-oxide suspended devices can be characterized by mechanical methods detecting their static displacement or eigenfrequency to correlate with the mechanical properties of their constituent materials,\cite{Davidovikj2020, Manca2021, Manca2023} paving the way towards the development of new functional micro-electro-mechanical devices.\cite{Manca2020, Lee2022a}

Mechanical resonators are among the most sensitive transducers and are employed in a variety of fields, from bio-molecules detection to quantum information processing. A critical parameter of mechanical resonators is their quality ($Q$) factor, which is a direct metric of their capability to sense variations of physical quantities such as mass or tension.\cite{Verbridge2006} $Q$ factor can be enhanced by lowering the relative energy loss per cycle towards the environment at the clamping points. This is possible by increasing the tension of the structures,\cite{Tsaturyan2017, Fedorov2019, Engelsen2021, Sementilli2022} or by realizing specific device geometries confining the mechanical modes within the resonator.\cite{Shin2022, Bereyhi2022a, Li2023} 
Designing complex oxide heterostructures by controlled growth allows for precise stress engineering,\cite{Zhang2001, MacManus-Driscoll2008, Cao2011, Mattoni2018} making these material suitable to realize tensile-stressed mechanical resonators with relatively high $Q$ factor values.\cite{Manca2022} The \ce{LaAlO3} (LAO) on \ce{SrTiO3} (STO) system is a good choice for such purpose as it shows a lattice mismatch of about 3\,\%. LAO thin films grown on STO substrates have large built-in tensile strain that can be partially relaxed by controlling the growth conditions.\cite{Sambri2020} 
Moreover, as LAO is a common substrate material for growing many oxides, LAO suspended structures can be ideal candidates for the realization of functional sensors or actuators.

\begin{figure*}
  \includegraphics[width=\linewidth]{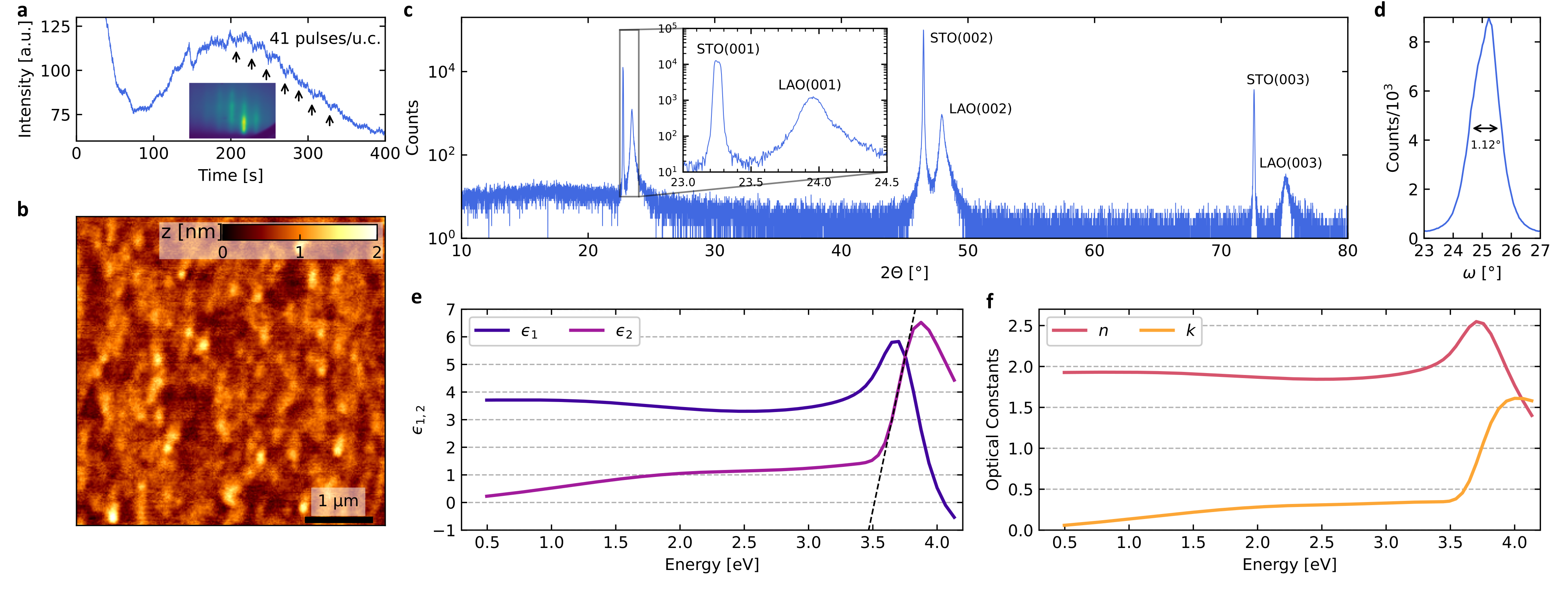}
  \caption{\label{fig:film}
    \ce{LaAlO3} thin films characterization.
    (a) Time evolution of the reflection high-energy electron diffraction (RHEED) intensity during the thin film growth. The inset shows the RHEED diffraction pattern at the end of the deposition.
    (b) Contact mode AFM topography image of the LAO surface.
    (c) X-rays $\Theta$--$2\Theta$ diffraction measurement of the LAO thin film. The inset shows a magnification of the STO and LAO (001) peaks.
    (d) $\omega$-scan (rocking curve) of the LAO(002) peak.
    (e) Coefficients of the complex dielectric response of the LAO thin films and
    (f) corresponding optical constants. The intersection at $\epsilon$=0 of the tangent to the onset of the $\epsilon_{\mathrm{2}}$ peak (black dashed line) gives an optical band-gap of about 3.4\,eV.
  }
\end{figure*}

In this work, we discuss the first realization of a  device consisting in a suspended high-Tc superconductor grown on top of a high $Q$ factor oxide mechanical resonator made of a single-crystal suspended LAO micro-bridge fabricated from a thin film deposited on top of a \ce{SrTiO3}(001) substrate.
Structural and optical properties of pristine LAO films are investigated by X-ray diffraction and spectroscopic ellipsometry.
Mechanical characterization of LAO micro-bridges, such as eigenfrequency and $Q$ factor length dependence, provides the in-plane stress and the intrinsic $Q$ value of the material.
Finally, we discuss the characteristics of YBCO grown on top of LAO micro-bridges by comparing their crystalline quality and transport properties with those of films grown on a STO(001) substrate.

\ce{LaAlO3} films with thickness of 100\,nm were deposited by pulsed laser deposition (PLD) on top of \ce{SrTiO3(001)} single crystal substrates from a 10$\times$10$\times$1\,mm$^3$ single crystal LAO target.
The growth temperature was 900\,\textcelsius, the background oxygen pressure was 1$\times$10$^{-4}$\,mbar, the laser fluency was 2\,J\,cm$^{-2}$, the laser repetition rate was 2\,Hz, and the target-substrate distance was about 3.5\,cm, resulting in a LAO growth rate of about 70\,nm/h.
The quality and deposition rate of the LAO film were monitored during the growth by reflection high-energy electron diffraction (RHEED). The time evolution of RHEED intensity is reported in Figure~\ref{fig:film}(a) and shows weak oscillations, corresponding to a growth rate of 41 pulses per unit cell, on top of a slow-varying non-monotonic background. Such time evolution has been already observed during the growth of low-stress LAO thin films.\cite{Sambri2020} The RHEED diffraction pattern at the end of the deposition (inset panel) shows striped features and absence of peaks related to 3D diffraction, as expected from thin films with low roughness. 
The LAO surface was inspected by atomic force microscopy and is reported in Fig.~\ref{fig:film}(b). It has a flat topography with a RMS roughness below 0.3\,nm over a 5$\times$5\,\textmu m$^2$ area.

Here, we focus on 100\,nm thick LAO films only because previous studies on other complex oxides MEMS showed that a good fabrication yield requires a film thickness of several tens of nanometers.\cite{Manca2023,Manca2022,Manca2021,Manca2020,Manca2019a}
However, for film thickness values above 25\,nm, the large epitaxial strain in LAO due to the STO substrate relaxes with the formation of cracks across the full height of the deposited films and may result in its delamination from the substrate.\cite{Sambri2020}
Consequently, to obtain LAO films of about 100\,nm and resistant to the fabrication of suspended structures we need to lower the built-in strain of the material. 
This was achieved by tuning the growth parameters, in particular oxygen partial pressure and temperature.

A certain amount of defects in our films is both expected and desirable and it is thus relevant to understand their impact on the material properties. We characterized the deposited LAO films by combining X-ray diffraction measurements (XRD) and spectroscopic ellipsometry measurements.
XRD measurements reported in Fig.~\ref{fig:film}(c) were performed in Bragg-Brentano configuration and show (00$l$) peaks without spurious phases. The inset panel is a close-up around the (001) peaks of the STO substrate and the LAO film. LAO is tensile strained due to the epitaxial mismatch with STO, with a $c$-axis of 3.81\,\AA. Considering the film thickness of 100\,nm, the diffraction peak is quite broad, suggesting a low crystal quality. This is also confirmed by the $\omega$-scan (rocking curve) reported in Fig~\ref{fig:film}(d) with a full width at half maximum of 1.12°, a relatively large value that signals the presence of crystallographic defects.
Crystal defects affect also the optical properties of LAO, which in bulk form is a transparent insulator with a band-gap of 5.6\,eV.\cite{Peacock2002} We thus checked a possible degradation of the LAO optical properties by spectroscopic ellipsometry.
Real ($\epsilon_{\mathrm{1}}$) and imaginary ($\epsilon_{\mathrm{2}}$) components of the dielectric response are reported in Fig.~\ref{fig:film}(d), while the corresponding optical constants $n$ and $k$ are reported in and Fig.~\ref{fig:film}(e).
The band-gap of our LAO thin films is 3.5\,eV, as obtained from the intersection at zero of the tangent to the maximal slope of $\varepsilon_2$ (black dashed line).
Its value is lower than the bulk one, but in good agreement with optical transmittance measurements performed on LAO/STO heterostructures under various growth conditions.\cite{Choi2012}
The presence of defects is particularly visible in the low-energy region, below the band-gap, of $\epsilon_{\mathrm{2}}$, which shows broadband absorption due to in-gap states.

\begin{figure}
  \includegraphics[width=\linewidth]{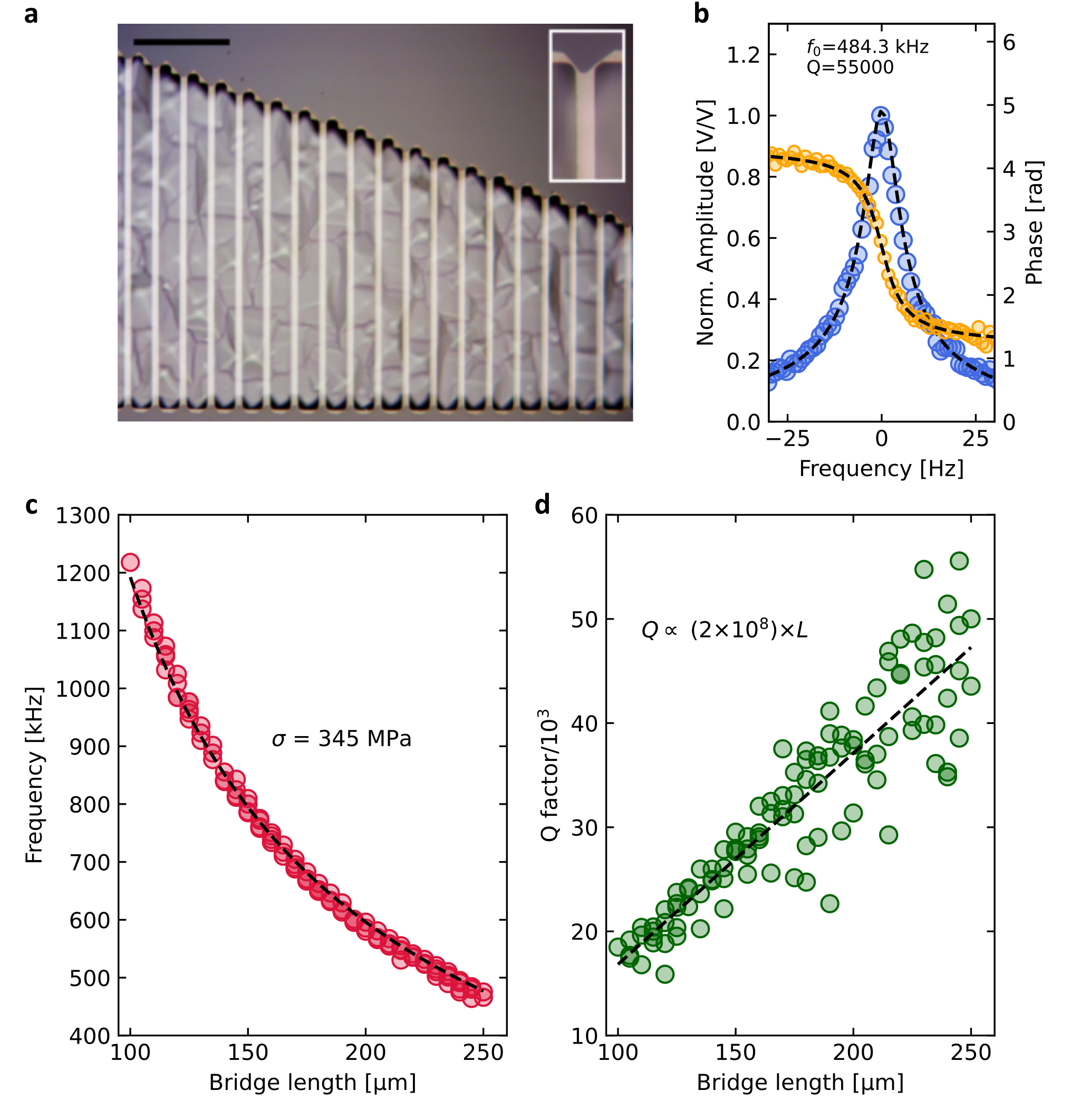}
  \caption{\label{fig:mec}
    Mechanical characterization of \ce{LaAlO3} micro-bridge resonators.
    (a) Optical micrograph of an harp array of LAO bridges. The inset shows a magnification of the clamping region. The black scale bar is 50\,\textmu m.
    (b)  Normalized magnitude and phase response of the first flexural mode of a 250\,\textmu m bridge. Dashed lines are best fit using Lorentzian response.
    (c) Length dependence of the first flexural mode eigenfrequency. The black dashed line is the best fit using Eq.\,(\ref{eq:string}) giving the in-plane stress ($\sigma$).
    (d) Length dependence of the $Q$ factor of the first flexural mode. The black dashed line is a linear fit.
  }
\end{figure}

Suspended micro-bridge structures were fabricated by standard UV lithography and Ar ions dry etching of the LAO films, followed by selective wet etching of the STO substrate in HF bath. Details of the fabrication process have already been reported elsewhere for other complex oxides grown on top of \ce{SrTiO3} substrates.\cite{Manca2019a, Manca2022, Manca2023}
An example of the fabricated devices is reported in Figure\,~\ref{fig:mec}(a), where suspended regions are light/yellow and clamped ones are dark/brown. The inset shows a magnification of the clamping region of a bridge, whose geometry is determined by the etching anisotropy of the STO substrate in HF.\cite{Plaza2021}
Micro-bridges length spans from 100 \textmu m to 250\,\textmu m with steps of 5\,\textmu m, while their width is about 4\,\textmu m. 
Mechanical spectral response of these resonators was measured in a custom setup featuring active temperature control and base pressure of 2$\times$10$^{-5}$\,mbar. Device motion is probed by optical lever detection scheme.
Mechanical excitation is provided by a AC-biased piezoelectric element glued by ceramic epoxy nearby the device.
Fig\,~\ref{fig:mec}(b) shows the resonance peak of the first flexural mode of a 250\,\textmu m-long LAO bridge.
Normalized magnitude (blue) and phase (orange) are both reported, together with the best fit of a Lorentzian response indicated as a black dashed line.
The resulting $Q$ factor of 55k is comparable to the highest values reported so far for complex oxides bridge resonators having similar size,\cite{Manca2022} likely related to the high tensile stress of the structures.
We evaluate the in-plane stress of LAO from the length dependence of the resonance frequency of the first flexural mode ($f_1$). For high-strain conditions, as expected from the large lattice mismatch between the LAO and the STO substrate, $f_1$ is given by the string resonator model\cite{Schmid2016}
\begin{equation}
  \label{eq:string}
  f_1 = \frac{1}{2L}\sqrt{\frac{\sigma}{\rho}},
\end{equation}
where $\rho$ is the LAO mass density (6070\,kg/m$^3$), $L$ is the bridge length, and $\sigma$ is the tensile stress.  Resonance frequency values measured for about 100 bridges having length from 100\,\textmu m to 250\,\textmu m are reported as red circles in Fig.~\ref{fig:mec}(c). Their best fit with Eq~(\ref{eq:string}) (black dashed line) provides an in-plane tensile stress of 345\,MPa. Such relatively high value, together with the good agreement between the data and the fitting function, confirm an high-stress condition of the LAO film.

In tension-dominated mechanical resonators, the intrinsic $Q$ factor ($Q_{\mathrm{int}}$) is enhanced due to dissipation-dilution mechanism,\cite{Sementilli2022} which is quantitatively represented by the dissipation dilution factor ($D$).
For a uniform string resonator having low aspect ratio, $D$ is given by the approximated expression
\begin{equation}
  \label{eq:diss_dil2}
  D = \sqrt{3} \sqrt{\frac{\sigma}{E}} \frac{L}{h},
\end{equation}
where $E$ the Young's modulus and $h$ is the thickness. This is in good agreement with the characteristics of our LAO resonators, which are uniform rectangular beams having $h/L < 10^{-3}$.
Moreover, if dissipation-dilution is the dominant mechanism determining the value of the quality factor of the resonator, $Q$ is proportional to the length of the bridges
\begin{equation}
  \label{eq:diss_q}
  Q = Q_{\mathrm{int}} \times D = Q_{\mathrm{int}}\frac{\sqrt{3}}{h} \sqrt{\frac{\sigma}{E}} \times L.
\end{equation}

In Fig.~\ref{fig:mec}(d),  we plot the length dependence of the $Q$ factor for the same set of resonators reported in Fig.~\ref{fig:mec}(c). The spread of the data is likely due to the cleanness of the sample after the fabrication process, which is still under optimization, but the large number of measured resonators allows us to identify a linear trend and a first order polynomial fit (black line) gives a slope value of about 2$\times$10$^8$\,m$^{-1}$.
By inverting Eq.~(\ref{eq:diss_q}) it is then possible to obtain the intrinsic $Q$ factor of LAO, which is a relevant material parameter to evaluate for MEMS applications.
Previous reports indicate that the Young's modulus of bulk LAO along the (001) is about 170\,GPa.\cite{Harrison2004, Carpenter2010b}
However, in 100\,nm-thick crystalline films of other oxides compounds, such as \ce{SrTiO3} and \ce{EuTiO3}, its has been observed an halving of $E$ with respect to the bulk values,\cite{Harbola2021,Manca2023} which greatly increases the uncertainty over the expected Young's modulus of our devices.
Taking the previous consideration into account, the calculated stress of $\sigma$=345\,MPa gives an in-plain strain in the 0.2--0.4\,\% range, which, notably, is in good agreement with previous measurements performed on tensile-strained \ce{WO3} and \ce{(La{,}Sr)MnO3} single crystal micro-bridge resonators.\cite{Manca2019a, Manca2022}
The resulting $Q_{\mathrm{int}}$ is then between 180 and 260, which is about 30 times lower than what recently measured in 100\,nm-thick \ce{Si3N4} resonators having same length-scale.\cite{Villanueva2014}
Such difference is a strong indication that the relatively high values of mechanical $Q$ factor measured on our LAO resonators could be further enhanced by improving material quality and, in case, the fabrication protocol.

\begin{figure}
  \includegraphics[width=\linewidth]{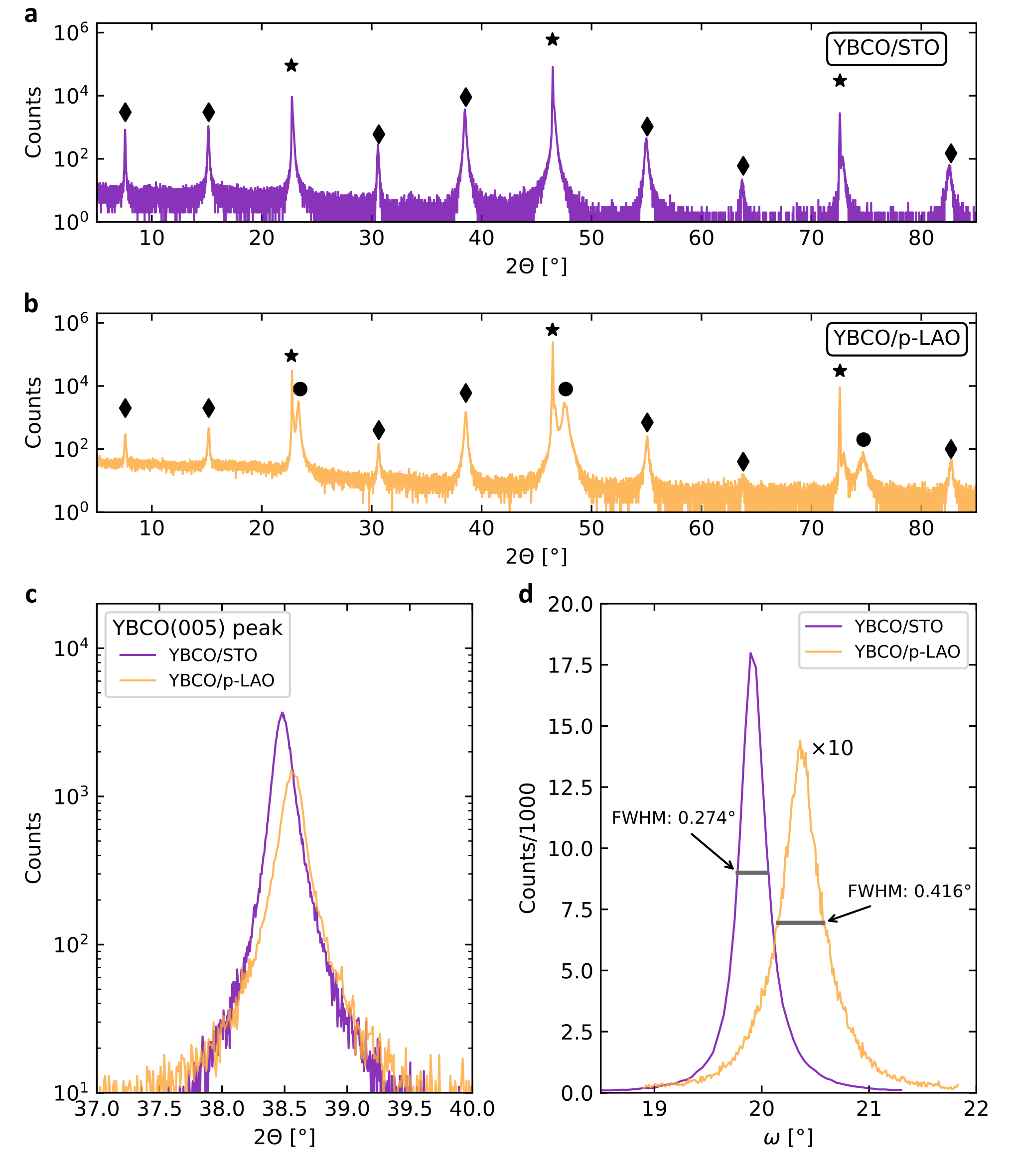}
  \caption{\label{fig:ybco}
    XRD analysis of \ce{YBCO} thin films grown on a STO(001) substrate (YBCO/STO) and a patterned LAO film (YBCO/p-LAO).
    (a, b) $\Theta$--$2\Theta$ scans of 100\,nm-thick YBCO films. Markers identify the material associated to each (00$l$) peak: STO ($\bigstar$), LAO ($\bullet$), and YBCO ($\blacklozenge$). YBCO(003)/(006)/(009) peaks are not marked because too close to the STO ones.
   (c, d) Comparison between the YBCO/STO and YBCO/p-LAO (005) diffraction peaks (c) and $\omega$-scans (d). The $\omega$-scan owning to the YBCO/p-LAO case is magnified ($\times$10) for better comparison. 
  }
\end{figure}

To demonstrate the potential of LAO suspended structures as templates for ex-situ hetero-epitaxial growth of complex oxides thin films, we selected YBCO, a high-Tc superconductor whose critical temperature is strongly affected by its crystal quality and strain.\cite{Mazini2022}
Figure~\ref{fig:ybco}(a) and (b) show the XRD $\Theta$--2$\Theta$ scans of two 100\,nm YBCO films deposited on a single-crystal STO(001) substrate (YBCO/STO) and on a patterned LAO film (YBCO/p-LAO).
For better clarity, in both the panels we marked the peaks owning to the same material: stars ($\bigstar$) for the STO substrate, ($\bullet$) for LAO, and diamonds ($\blacklozenge$) for the YBCO. YBCO(003)/(006)/(009) peaks are too close to the STO ones and were not marked to avoid crowding.
For better comparison of the two diffraction patterns, we show in Fig~\ref{fig:ybco}(c) the (005) peak of the YBCO deposited the STO substrate and on the patterned LAO film. They have an almost identical shape, but the one belonging to the YBCO/p-LAO shows a +0.09\,° shift, corresponding to a -0.19\,\% variation in the out-of-plane direction.
The YBCO $c$-axis is thus 11.68\,\AA\ for the film deposited on the STO substrate, identical to its bulk value, while 11.66\,\AA\ for the film deposited on the patterned LAO.
We found a more marked difference between the two cases when comparing their rocking curves. $\omega$-scans measurements of the YBCO(005) peaks are reported in Fig.~\ref{fig:ybco}(d), where we also show their full width at half maximum (FWHM), which is an indication of crystalline disorder. A larger value is found for the YBCO/p-LAO case, signalling lower crystal quality. This is expected, since the growth on STO(001) can be considered an ideal case, while it is relevant to note that the peak width of YBCO on patterned LAO is much lower than the one of the underlying LAO film, which is reported in Fig.~\ref{fig:film}(d).

\begin{figure}[b]
  \includegraphics[width=\linewidth]{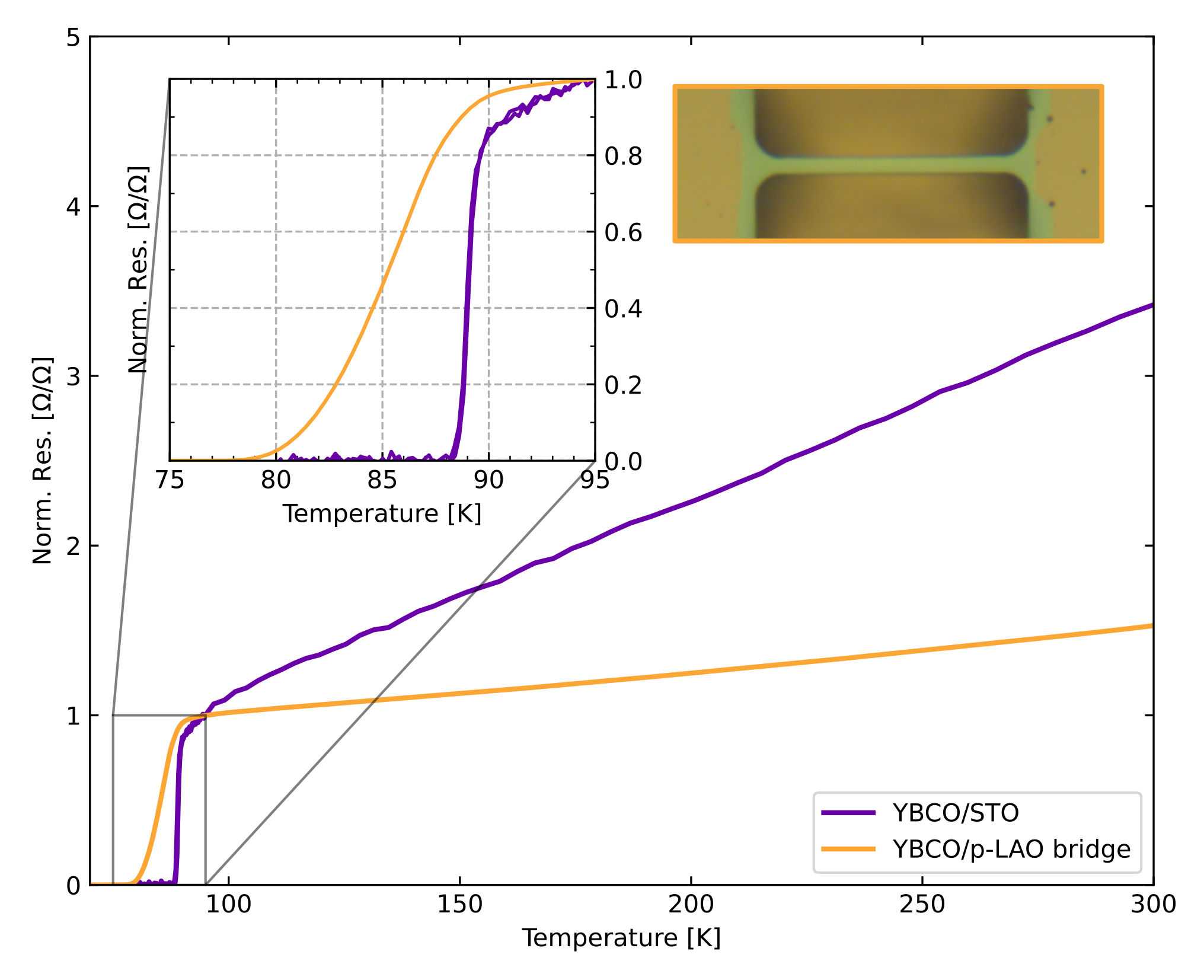}
  \caption{\label{fig:rt}
     Temperature dependence of the normalized electrical resistance of YBCO deposited on a STO substrate (purple) and a LAO micro-bridge (orange) as those shown in the optical micrograph. The close-up inset shows the resistance evolution around the superconducting transition temperature.
  }
\end{figure}

Although the diffraction pattern of YBCO on LAO points towards high crystal quality, showing no spurious phases and an almost identical amplitude of the peaks with respect to the film grown on top of the STO substrate, we can not obtain information about the film deposited on the suspended regions as their volume is negligible with respect to the one of the pad regions.
To directly probe the characteristics of the YBCO on suspended LAO, we thus performed four-probe electrical transport measurements across a micro-bridge.
The temperature dependence of the electrical resistance was measured from 300\,K down to 70\,K and is reported in Figure~\ref{fig:rt} (orange line).
The measured bridge has a width of 4\,\textmu m and a length of 50\,\textmu m and its optical micrograph is included as inset of the same figure panel.
The electrical connections were realized by ultrasonic bonding wires on the 200$\times$200\,\textmu m$^2$ pads located at the two ends of a bridge.
The YBCO film is found to be superconducting even across the narrow freestanding membrane. The onset of a superconducting transition is at $\mathrm{T_{C,ON}}$=90\,K, reaching zero resistance state at around $\mathrm{T_{C,0}}$=78\,K. The transition is very broad as compared with the YBCO film deposited on the STO substrate under similar conditions (purple line), where the width of the corresponding transition is less than 1\,K. This broadening may be caused by various factors, such as crystalline quality, oxygen non-stoichiometry, or mechanical strain. We note that a long post annealing in oxygen atmosphere did not change the superconducting properties. The XRD data also indicate that the $c$-axis lattice parameter of the YBCO/p-LAO film corresponds to the optimally doped YBCO.\cite{Benzi2004} This suggests that the broadening of transition is probably caused by lower crystalline quality of the YBCO grown on the LAO, as indicated by the broader rocking curve of Fig.~\ref{fig:ybco}(d).
These results are very promising for studying the fundamental properties of YBCO and for the fabrication of sensors based on suspended YBCO structures, such as kinetic inductance bolometers, magnetic field sensors, or superconducting opto-mechanical resonators.\cite{Chakrabarty2019, Maspero2021}

In conclusion, we demonstrated the fabrication of single crystal LAO suspended bridges as templates for ex-situ growth of other functional complex oxides.
Their mechanical characterization provided an in-plane stress of almost 350\,MPa, mechanical $Q$ factors of several tens of thousands, and an intrinsic $Q$ factor for LAO of about 250. This latter value is rather low if compared with Si compounds. More experiments are needed to clarify the intrinsic limits of this material – and other similar oxides - in terms of mechanical losses, and if they can be improved (and how much) by controlling the growth conditions and the fabrication processes.
Single crystal YBCO thin films deposited on top of LAO MEMS show high crystal quality and superconducting properties. Further optimization process is required if compared with optimally-doped film grown on STO(001) substrates.
We demonstrated the integration of a high-Tc superconductor on top of a transparent and tensile-strained oxide micro-bridge. Potential applications of the presented device include bolometers, thanks to its weak thermal coupling, magnetometers,  and opto-mechanical experiments related to fundamental studies of the superconducting transition, in particular in its mixed state.

\section*{Acknowledgements}

This work was carried out under the OXiNEMS project (\href{www.oxinems.eu}{www.oxinems.eu}). This project has received funding from the European Union’s Horizon 2020 research and innovation programme under Grant Agreement No. 828784. We acknowledge financial support from the Universit\`a di Genova through the ``Fondi di Ricerca di Ateneo'' (FRA). We also acknowledge support from the Swedish infrastructure for micro- and nano-fabrication - MyFab.

\section*{Open Data}

The numerical data shown in figures of the manuscript 
can be donwloaded from the Zenodo online repository: ~\href{http://dx.doi.org/10.5281/zenodo.10521695}{http://dx.doi.org/10.5281/zenodo.10521695}

\bibliography{Library.bib}

\begin{thebibliography}{40}%
\makeatletter
\providecommand \@ifxundefined [1]{%
 \@ifx{#1\undefined}
}%
\providecommand \@ifnum [1]{%
 \ifnum #1\expandafter \@firstoftwo
 \else \expandafter \@secondoftwo
 \fi
}%
\providecommand \@ifx [1]{%
 \ifx #1\expandafter \@firstoftwo
 \else \expandafter \@secondoftwo
 \fi
}%
\providecommand \natexlab [1]{#1}%
\providecommand \enquote  [1]{``#1''}%
\providecommand \bibnamefont  [1]{#1}%
\providecommand \bibfnamefont [1]{#1}%
\providecommand \citenamefont [1]{#1}%
\providecommand \href@noop [0]{\@secondoftwo}%
\providecommand \href [0]{\begingroup \@sanitize@url \@href}%
\providecommand \@href[1]{\@@startlink{#1}\@@href}%
\providecommand \@@href[1]{\endgroup#1\@@endlink}%
\providecommand \@sanitize@url [0]{\catcode `\\12\catcode `\$12\catcode
  `\&12\catcode `\#12\catcode `\^12\catcode `\_12\catcode `\%12\relax}%
\providecommand \@@startlink[1]{}%
\providecommand \@@endlink[0]{}%
\providecommand \url  [0]{\begingroup\@sanitize@url \@url }%
\providecommand \@url [1]{\endgroup\@href {#1}{\urlprefix }}%
\providecommand \urlprefix  [0]{URL }%
\providecommand \Eprint [0]{\href }%
\providecommand \doibase [0]{https://doi.org/}%
\providecommand \selectlanguage [0]{\@gobble}%
\providecommand \bibinfo  [0]{\@secondoftwo}%
\providecommand \bibfield  [0]{\@secondoftwo}%
\providecommand \translation [1]{[#1]}%
\providecommand \BibitemOpen [0]{}%
\providecommand \bibitemStop [0]{}%
\providecommand \bibitemNoStop [0]{.\EOS\space}%
\providecommand \EOS [0]{\spacefactor3000\relax}%
\providecommand \BibitemShut  [1]{\csname bibitem#1\endcsname}%
\let\auto@bib@innerbib\@empty
\bibitem [{\citenamefont {Ramesh}\ and\ \citenamefont
  {Schlom}(2019)}]{Ramesh2019}%
  \BibitemOpen
  \bibfield  {author} {\bibinfo {author} {\bibfnamefont {R.}~\bibnamefont
  {Ramesh}}\ and\ \bibinfo {author} {\bibfnamefont {D.~G.}\ \bibnamefont
  {Schlom}},\ }\bibfield  {title} {\bibinfo {title} {Creating emergent
  phenomena in oxide superlattices},\ }\href
  {https://doi.org/10.1038/s41578-019-0095-2} {\bibfield  {journal} {\bibinfo
  {journal} {Nat Rev Mater}\ }\textbf {\bibinfo {volume} {4}},\ \bibinfo
  {pages} {257} (\bibinfo {year} {2019})}\BibitemShut {NoStop}%
\bibitem [{\citenamefont {Zubko}\ \emph {et~al.}(2011)\citenamefont {Zubko},
  \citenamefont {Gariglio}, \citenamefont {Gabay}, \citenamefont {Ghosez},\
  and\ \citenamefont {Triscone}}]{Zubko2011}%
  \BibitemOpen
  \bibfield  {author} {\bibinfo {author} {\bibfnamefont {P.}~\bibnamefont
  {Zubko}}, \bibinfo {author} {\bibfnamefont {S.}~\bibnamefont {Gariglio}},
  \bibinfo {author} {\bibfnamefont {M.}~\bibnamefont {Gabay}}, \bibinfo
  {author} {\bibfnamefont {P.}~\bibnamefont {Ghosez}},\ and\ \bibinfo {author}
  {\bibfnamefont {J.-M.}\ \bibnamefont {Triscone}},\ }\bibfield  {title}
  {\bibinfo {title} {Interface {{Physics}} in {{Complex Oxide
  Heterostructures}}},\ }\href
  {https://doi.org/10.1146/annurev-conmatphys-062910-140445} {\bibfield
  {journal} {\bibinfo  {journal} {Annu. Rev. Condens. Matter Phys.}\ }\textbf
  {\bibinfo {volume} {2}},\ \bibinfo {pages} {141} (\bibinfo {year}
  {2011})}\BibitemShut {NoStop}%
\bibitem [{\citenamefont {Pai}\ \emph {et~al.}(2018)\citenamefont {Pai},
  \citenamefont {Lee}, \citenamefont {Lee}, \citenamefont {Annadi},
  \citenamefont {Cheng}, \citenamefont {Lu}, \citenamefont {Tomczyk},
  \citenamefont {Huang}, \citenamefont {Eom}, \citenamefont {Irvin},\ and\
  \citenamefont {Levy}}]{Pai2018a}%
  \BibitemOpen
  \bibfield  {author} {\bibinfo {author} {\bibfnamefont {Y.-Y.}\ \bibnamefont
  {Pai}}, \bibinfo {author} {\bibfnamefont {H.}~\bibnamefont {Lee}}, \bibinfo
  {author} {\bibfnamefont {J.-W.}\ \bibnamefont {Lee}}, \bibinfo {author}
  {\bibfnamefont {A.}~\bibnamefont {Annadi}}, \bibinfo {author} {\bibfnamefont
  {G.}~\bibnamefont {Cheng}}, \bibinfo {author} {\bibfnamefont
  {S.}~\bibnamefont {Lu}}, \bibinfo {author} {\bibfnamefont {M.}~\bibnamefont
  {Tomczyk}}, \bibinfo {author} {\bibfnamefont {M.}~\bibnamefont {Huang}},
  \bibinfo {author} {\bibfnamefont {C.-B.}\ \bibnamefont {Eom}}, \bibinfo
  {author} {\bibfnamefont {P.}~\bibnamefont {Irvin}},\ and\ \bibinfo {author}
  {\bibfnamefont {J.}~\bibnamefont {Levy}},\ }\bibfield  {title} {\bibinfo
  {title} {One-{{Dimensional Nature}} of {{Superconductivity}} at the
  {{LaAlO3}}/{{SrTiO3 Interface}}},\ }\href
  {https://doi.org/10.1103/PhysRevLett.120.147001} {\bibfield  {journal}
  {\bibinfo  {journal} {Phys. Rev. Lett.}\ }\textbf {\bibinfo {volume} {120}},\
  \bibinfo {pages} {147001} (\bibinfo {year} {2018})}\BibitemShut {NoStop}%
\bibitem [{\citenamefont {Christensen}\ \emph {et~al.}(2019)\citenamefont
  {Christensen}, \citenamefont {Trier}, \citenamefont {Niu}, \citenamefont
  {Gan}, \citenamefont {Zhang}, \citenamefont {Jespersen}, \citenamefont
  {Chen},\ and\ \citenamefont {Pryds}}]{Christensen2019}%
  \BibitemOpen
  \bibfield  {author} {\bibinfo {author} {\bibfnamefont {D.~V.}\ \bibnamefont
  {Christensen}}, \bibinfo {author} {\bibfnamefont {F.}~\bibnamefont {Trier}},
  \bibinfo {author} {\bibfnamefont {W.}~\bibnamefont {Niu}}, \bibinfo {author}
  {\bibfnamefont {Y.}~\bibnamefont {Gan}}, \bibinfo {author} {\bibfnamefont
  {Y.}~\bibnamefont {Zhang}}, \bibinfo {author} {\bibfnamefont {T.~S.}\
  \bibnamefont {Jespersen}}, \bibinfo {author} {\bibfnamefont {Y.}~\bibnamefont
  {Chen}},\ and\ \bibinfo {author} {\bibfnamefont {N.}~\bibnamefont {Pryds}},\
  }\bibfield  {title} {\bibinfo {title} {Stimulating {{Oxide
  Heterostructures}}: {{A Review}} on {{Controlling
  SrTiO}}{\textsubscript{3}}-{{Based Heterointerfaces}} with {{External
  Stimuli}}},\ }\href {https://doi.org/10.1002/admi.201900772} {\bibfield
  {journal} {\bibinfo  {journal} {Adv. Mater. Interfaces}\ }\textbf {\bibinfo
  {volume} {6}},\ \bibinfo {pages} {1900772} (\bibinfo {year}
  {2019})}\BibitemShut {NoStop}%
\bibitem [{\citenamefont {Stornaiuolo}\ \emph {et~al.}(2016)\citenamefont
  {Stornaiuolo}, \citenamefont {Cantoni}, \citenamefont {De~Luca},
  \citenamefont {Di~Capua}, \citenamefont {Di.~Gennaro}, \citenamefont
  {Ghiringhelli}, \citenamefont {Jouault}, \citenamefont {Marr{\`e}},
  \citenamefont {Massarotti}, \citenamefont {Miletto~Granozio}, \citenamefont
  {Pallecchi}, \citenamefont {Piamonteze}, \citenamefont {Rusponi},
  \citenamefont {Tafuri},\ and\ \citenamefont {Salluzzo}}]{Stornaiuolo2016}%
  \BibitemOpen
  \bibfield  {author} {\bibinfo {author} {\bibfnamefont {D.}~\bibnamefont
  {Stornaiuolo}}, \bibinfo {author} {\bibfnamefont {C.}~\bibnamefont
  {Cantoni}}, \bibinfo {author} {\bibfnamefont {G.~M.}\ \bibnamefont
  {De~Luca}}, \bibinfo {author} {\bibfnamefont {R.}~\bibnamefont {Di~Capua}},
  \bibinfo {author} {\bibfnamefont {E.}~\bibnamefont {Di.~Gennaro}}, \bibinfo
  {author} {\bibfnamefont {G.}~\bibnamefont {Ghiringhelli}}, \bibinfo {author}
  {\bibfnamefont {B.}~\bibnamefont {Jouault}}, \bibinfo {author} {\bibfnamefont
  {D.}~\bibnamefont {Marr{\`e}}}, \bibinfo {author} {\bibfnamefont
  {D.}~\bibnamefont {Massarotti}}, \bibinfo {author} {\bibfnamefont
  {F.}~\bibnamefont {Miletto~Granozio}}, \bibinfo {author} {\bibfnamefont
  {I.}~\bibnamefont {Pallecchi}}, \bibinfo {author} {\bibfnamefont
  {C.}~\bibnamefont {Piamonteze}}, \bibinfo {author} {\bibfnamefont
  {S.}~\bibnamefont {Rusponi}}, \bibinfo {author} {\bibfnamefont
  {F.}~\bibnamefont {Tafuri}},\ and\ \bibinfo {author} {\bibfnamefont
  {M.}~\bibnamefont {Salluzzo}},\ }\bibfield  {title} {\bibinfo {title}
  {Tunable spin polarization and superconductivity in engineered oxide
  interfaces},\ }\href {https://doi.org/10.1038/nmat4491} {\bibfield  {journal}
  {\bibinfo  {journal} {Nature Mater}\ }\textbf {\bibinfo {volume} {15}},\
  \bibinfo {pages} {278} (\bibinfo {year} {2016})}\BibitemShut {NoStop}%
\bibitem [{\citenamefont {Br{\'e}hin}\ \emph {et~al.}(2023)\citenamefont
  {Br{\'e}hin}, \citenamefont {Chen}, \citenamefont {D'Antuono}, \citenamefont
  {Varotto}, \citenamefont {Stornaiuolo}, \citenamefont {Piamonteze},
  \citenamefont {Varignon}, \citenamefont {Salluzzo},\ and\ \citenamefont
  {Bibes}}]{Brehin2023}%
  \BibitemOpen
  \bibfield  {author} {\bibinfo {author} {\bibfnamefont {J.}~\bibnamefont
  {Br{\'e}hin}}, \bibinfo {author} {\bibfnamefont {Y.}~\bibnamefont {Chen}},
  \bibinfo {author} {\bibfnamefont {M.}~\bibnamefont {D'Antuono}}, \bibinfo
  {author} {\bibfnamefont {S.}~\bibnamefont {Varotto}}, \bibinfo {author}
  {\bibfnamefont {D.}~\bibnamefont {Stornaiuolo}}, \bibinfo {author}
  {\bibfnamefont {C.}~\bibnamefont {Piamonteze}}, \bibinfo {author}
  {\bibfnamefont {J.}~\bibnamefont {Varignon}}, \bibinfo {author}
  {\bibfnamefont {M.}~\bibnamefont {Salluzzo}},\ and\ \bibinfo {author}
  {\bibfnamefont {M.}~\bibnamefont {Bibes}},\ }\bibfield  {title} {\bibinfo
  {title} {Coexistence and coupling of ferroelectricity and magnetism in an
  oxide two-dimensional electron gas},\ }\bibfield  {journal} {\bibinfo
  {journal} {Nat. Phys.}\ }\href {https://doi.org/10.1038/s41567-023-01983-y}
  {10.1038/s41567-023-01983-y} (\bibinfo {year} {2023})\BibitemShut {NoStop}%
\bibitem [{\citenamefont {Biasotti}\ \emph {et~al.}(2010)\citenamefont
  {Biasotti}, \citenamefont {Pellegrino}, \citenamefont {Bellingeri},
  \citenamefont {Manca}, \citenamefont {Siri},\ and\ \citenamefont
  {Marr{\`e}}}]{Biasotti2010}%
  \BibitemOpen
  \bibfield  {author} {\bibinfo {author} {\bibfnamefont {M.}~\bibnamefont
  {Biasotti}}, \bibinfo {author} {\bibfnamefont {L.}~\bibnamefont
  {Pellegrino}}, \bibinfo {author} {\bibfnamefont {E.}~\bibnamefont
  {Bellingeri}}, \bibinfo {author} {\bibfnamefont {N.}~\bibnamefont {Manca}},
  \bibinfo {author} {\bibfnamefont {A.~S.}\ \bibnamefont {Siri}},\ and\
  \bibinfo {author} {\bibfnamefont {D.}~\bibnamefont {Marr{\`e}}},\ }\bibfield
  {title} {\bibinfo {title} {Strain response of
  {{La}}{\textsubscript{0.7}}{{Sr}}{\textsubscript{0.3}}{{CoO}}{\textsubscript{3}}
  epitaxial thin films probed by {{SrTiO}}{\textsubscript{3}} crystalline
  microcantilevers},\ }\href {https://doi.org/10.1063/1.3519478} {\bibfield
  {journal} {\bibinfo  {journal} {Appl. Phys. Lett.}\ }\textbf {\bibinfo
  {volume} {97}},\ \bibinfo {pages} {223503} (\bibinfo {year}
  {2010})}\BibitemShut {NoStop}%
\bibitem [{\citenamefont {Pellegrino}\ \emph {et~al.}(2012)\citenamefont
  {Pellegrino}, \citenamefont {Manca}, \citenamefont {Kanki}, \citenamefont
  {Tanaka}, \citenamefont {Biasotti}, \citenamefont {Bellingeri}, \citenamefont
  {Siri},\ and\ \citenamefont {Marr{\'e}}}]{Pellegrino2012}%
  \BibitemOpen
  \bibfield  {author} {\bibinfo {author} {\bibfnamefont {L.}~\bibnamefont
  {Pellegrino}}, \bibinfo {author} {\bibfnamefont {N.}~\bibnamefont {Manca}},
  \bibinfo {author} {\bibfnamefont {T.}~\bibnamefont {Kanki}}, \bibinfo
  {author} {\bibfnamefont {H.}~\bibnamefont {Tanaka}}, \bibinfo {author}
  {\bibfnamefont {M.}~\bibnamefont {Biasotti}}, \bibinfo {author}
  {\bibfnamefont {E.}~\bibnamefont {Bellingeri}}, \bibinfo {author}
  {\bibfnamefont {A.~S.}\ \bibnamefont {Siri}},\ and\ \bibinfo {author}
  {\bibfnamefont {D.}~\bibnamefont {Marr{\'e}}},\ }\bibfield  {title} {\bibinfo
  {title} {Multistate {{Memory Devices Based}} on {{Free-standing
  VO}}{\textsubscript{2}}/{{TiO}}{\textsubscript{2}} {{Microstructures Driven}}
  by {{Joule Self-Heating}}},\ }\href {https://doi.org/10.1002/adma.201104669}
  {\bibfield  {journal} {\bibinfo  {journal} {Adv. Mater.}\ }\textbf {\bibinfo
  {volume} {24}},\ \bibinfo {pages} {2929} (\bibinfo {year}
  {2012})}\BibitemShut {NoStop}%
\bibitem [{\citenamefont {Davidovikj}\ \emph {et~al.}(2020)\citenamefont
  {Davidovikj}, \citenamefont {Groenendijk}, \citenamefont {Monteiro},
  \citenamefont {Dijkhoff}, \citenamefont {Afanasiev}, \citenamefont {{\v
  S}i{\v s}kins}, \citenamefont {Lee}, \citenamefont {Huang}, \citenamefont
  {{van Heumen}}, \citenamefont {{van der Zant}}, \citenamefont {Caviglia},\
  and\ \citenamefont {Steeneken}}]{Davidovikj2020}%
  \BibitemOpen
  \bibfield  {author} {\bibinfo {author} {\bibfnamefont {D.}~\bibnamefont
  {Davidovikj}}, \bibinfo {author} {\bibfnamefont {D.~J.}\ \bibnamefont
  {Groenendijk}}, \bibinfo {author} {\bibfnamefont {A.~M. R. V.~L.}\
  \bibnamefont {Monteiro}}, \bibinfo {author} {\bibfnamefont {A.}~\bibnamefont
  {Dijkhoff}}, \bibinfo {author} {\bibfnamefont {D.}~\bibnamefont {Afanasiev}},
  \bibinfo {author} {\bibfnamefont {M.}~\bibnamefont {{\v S}i{\v s}kins}},
  \bibinfo {author} {\bibfnamefont {M.}~\bibnamefont {Lee}}, \bibinfo {author}
  {\bibfnamefont {Y.}~\bibnamefont {Huang}}, \bibinfo {author} {\bibfnamefont
  {E.}~\bibnamefont {{van Heumen}}}, \bibinfo {author} {\bibfnamefont
  {H.~S.~J.}\ \bibnamefont {{van der Zant}}}, \bibinfo {author} {\bibfnamefont
  {A.~D.}\ \bibnamefont {Caviglia}},\ and\ \bibinfo {author} {\bibfnamefont
  {P.~G.}\ \bibnamefont {Steeneken}},\ }\bibfield  {title} {\bibinfo {title}
  {Ultrathin complex oxide nanomechanical resonators},\ }\href
  {https://doi.org/10.1038/s42005-020-00433-y} {\bibfield  {journal} {\bibinfo
  {journal} {Commun Phys}\ }\textbf {\bibinfo {volume} {3}},\ \bibinfo {pages}
  {163} (\bibinfo {year} {2020})}\BibitemShut {NoStop}%
\bibitem [{\citenamefont {Manca}\ \emph {et~al.}(2021)\citenamefont {Manca},
  \citenamefont {Kanki}, \citenamefont {Endo}, \citenamefont {Ragucci},
  \citenamefont {Pellegrino},\ and\ \citenamefont {Marr{\'e}}}]{Manca2021}%
  \BibitemOpen
  \bibfield  {author} {\bibinfo {author} {\bibfnamefont {N.}~\bibnamefont
  {Manca}}, \bibinfo {author} {\bibfnamefont {T.}~\bibnamefont {Kanki}},
  \bibinfo {author} {\bibfnamefont {F.}~\bibnamefont {Endo}}, \bibinfo {author}
  {\bibfnamefont {E.}~\bibnamefont {Ragucci}}, \bibinfo {author} {\bibfnamefont
  {L.}~\bibnamefont {Pellegrino}},\ and\ \bibinfo {author} {\bibfnamefont
  {D.}~\bibnamefont {Marr{\'e}}},\ }\bibfield  {title} {\bibinfo {title}
  {Anisotropic {{Temperature-Driven Strain Dynamics}} in
  {{VO}}{\textsubscript{2}} {{Solid-State Microactuators}}},\ }\href
  {https://doi.org/10.1021/acsaelm.0c00776} {\bibfield  {journal} {\bibinfo
  {journal} {ACS Appl. Electron. Mater.}\ }\textbf {\bibinfo {volume} {3}},\
  \bibinfo {pages} {211} (\bibinfo {year} {2021})}\BibitemShut {NoStop}%
\bibitem [{\citenamefont {Manca}\ \emph {et~al.}(2023)\citenamefont {Manca},
  \citenamefont {Tarsi}, \citenamefont {Kalaboukhov}, \citenamefont {Bisio},
  \citenamefont {Caglieris}, \citenamefont {Lombardi}, \citenamefont
  {Marr{\'e}},\ and\ \citenamefont {Pellegrino}}]{Manca2023}%
  \BibitemOpen
  \bibfield  {author} {\bibinfo {author} {\bibfnamefont {N.}~\bibnamefont
  {Manca}}, \bibinfo {author} {\bibfnamefont {G.}~\bibnamefont {Tarsi}},
  \bibinfo {author} {\bibfnamefont {A.}~\bibnamefont {Kalaboukhov}}, \bibinfo
  {author} {\bibfnamefont {F.}~\bibnamefont {Bisio}}, \bibinfo {author}
  {\bibfnamefont {F.}~\bibnamefont {Caglieris}}, \bibinfo {author}
  {\bibfnamefont {F.}~\bibnamefont {Lombardi}}, \bibinfo {author}
  {\bibfnamefont {D.}~\bibnamefont {Marr{\'e}}},\ and\ \bibinfo {author}
  {\bibfnamefont {L.}~\bibnamefont {Pellegrino}},\ }\bibfield  {title}
  {\bibinfo {title} {Strain, {{Young}}'s modulus, and structural transition of
  {{EuTiO3}} thin films probed by micro-mechanical methods},\ }\bibfield
  {journal} {\bibinfo  {journal} {APL Mater.}\ }\textbf {\bibinfo {volume}
  {11}},\ \href {https://doi.org/10.1063/5.0166762} {10.1063/5.0166762}
  (\bibinfo {year} {2023})\BibitemShut {NoStop}%
\bibitem [{\citenamefont {Manca}\ \emph {et~al.}(2020)\citenamefont {Manca},
  \citenamefont {Kanki}, \citenamefont {Endo}, \citenamefont {Marr{\'e}},\ and\
  \citenamefont {Pellegrino}}]{Manca2020}%
  \BibitemOpen
  \bibfield  {author} {\bibinfo {author} {\bibfnamefont {N.}~\bibnamefont
  {Manca}}, \bibinfo {author} {\bibfnamefont {T.}~\bibnamefont {Kanki}},
  \bibinfo {author} {\bibfnamefont {F.}~\bibnamefont {Endo}}, \bibinfo {author}
  {\bibfnamefont {D.}~\bibnamefont {Marr{\'e}}},\ and\ \bibinfo {author}
  {\bibfnamefont {L.}~\bibnamefont {Pellegrino}},\ }\bibfield  {title}
  {\bibinfo {title} {Planar {{Nanoactuators Based}} on
  {{VO}}{\textsubscript{2}} {{Phase Transition}}},\ }\href
  {https://doi.org/10.1021/acs.nanolett.0c02638} {\bibfield  {journal}
  {\bibinfo  {journal} {Nano Lett.}\ }\textbf {\bibinfo {volume} {20}},\
  \bibinfo {pages} {7251} (\bibinfo {year} {2020})}\BibitemShut {NoStop}%
\bibitem [{\citenamefont {Lee}\ \emph {et~al.}(2022)\citenamefont {Lee},
  \citenamefont {Renshof}, \citenamefont {{van Zeggeren}}, \citenamefont
  {Houmes}, \citenamefont {Lesne}, \citenamefont {{\v S}i{\v s}kins},
  \citenamefont {{van Thiel}}, \citenamefont {Guis}, \citenamefont {{van
  Blankenstein}}, \citenamefont {Verbiest}, \citenamefont {Caviglia},
  \citenamefont {{van der Zant}},\ and\ \citenamefont {Steeneken}}]{Lee2022a}%
  \BibitemOpen
  \bibfield  {author} {\bibinfo {author} {\bibfnamefont {M.}~\bibnamefont
  {Lee}}, \bibinfo {author} {\bibfnamefont {J.~R.}\ \bibnamefont {Renshof}},
  \bibinfo {author} {\bibfnamefont {K.~J.}\ \bibnamefont {{van Zeggeren}}},
  \bibinfo {author} {\bibfnamefont {M.~J.~A.}\ \bibnamefont {Houmes}}, \bibinfo
  {author} {\bibfnamefont {E.}~\bibnamefont {Lesne}}, \bibinfo {author}
  {\bibfnamefont {M.}~\bibnamefont {{\v S}i{\v s}kins}}, \bibinfo {author}
  {\bibfnamefont {T.~C.}\ \bibnamefont {{van Thiel}}}, \bibinfo {author}
  {\bibfnamefont {R.~H.}\ \bibnamefont {Guis}}, \bibinfo {author}
  {\bibfnamefont {M.~R.}\ \bibnamefont {{van Blankenstein}}}, \bibinfo {author}
  {\bibfnamefont {G.~J.}\ \bibnamefont {Verbiest}}, \bibinfo {author}
  {\bibfnamefont {A.~D.}\ \bibnamefont {Caviglia}}, \bibinfo {author}
  {\bibfnamefont {H.~S.~J.}\ \bibnamefont {{van der Zant}}},\ and\ \bibinfo
  {author} {\bibfnamefont {P.~G.}\ \bibnamefont {Steeneken}},\ }\bibfield
  {title} {\bibinfo {title} {Ultrathin {{Piezoelectric Resonators Based}} on
  {{Graphene}} and {{Free}}-{{Standing Single}}-{{Crystal
  BaTiO}}{\textsubscript{3}}},\ }\href {https://doi.org/10.1002/adma.202204630}
  {\bibfield  {journal} {\bibinfo  {journal} {Advanced Materials}\ }\textbf
  {\bibinfo {volume} {34}},\ \bibinfo {pages} {2204630} (\bibinfo {year}
  {2022})}\BibitemShut {NoStop}%
\bibitem [{\citenamefont {Verbridge}\ \emph {et~al.}(2006)\citenamefont
  {Verbridge}, \citenamefont {Parpia}, \citenamefont {Reichenbach},
  \citenamefont {Bellan},\ and\ \citenamefont {Craighead}}]{Verbridge2006}%
  \BibitemOpen
  \bibfield  {author} {\bibinfo {author} {\bibfnamefont {S.~S.}\ \bibnamefont
  {Verbridge}}, \bibinfo {author} {\bibfnamefont {J.~M.}\ \bibnamefont
  {Parpia}}, \bibinfo {author} {\bibfnamefont {R.~B.}\ \bibnamefont
  {Reichenbach}}, \bibinfo {author} {\bibfnamefont {L.~M.}\ \bibnamefont
  {Bellan}},\ and\ \bibinfo {author} {\bibfnamefont {H.~G.}\ \bibnamefont
  {Craighead}},\ }\bibfield  {title} {\bibinfo {title} {High quality factor
  resonance at room temperature with nanostrings under high tensile stress},\
  }\href {https://doi.org/10.1063/1.2204829} {\bibfield  {journal} {\bibinfo
  {journal} {Journal of Applied Physics}\ }\textbf {\bibinfo {volume} {99}},\
  \bibinfo {pages} {124304} (\bibinfo {year} {2006})}\BibitemShut {NoStop}%
\bibitem [{\citenamefont {Tsaturyan}\ \emph {et~al.}(2017)\citenamefont
  {Tsaturyan}, \citenamefont {Barg}, \citenamefont {Polzik},\ and\
  \citenamefont {Schliesser}}]{Tsaturyan2017}%
  \BibitemOpen
  \bibfield  {author} {\bibinfo {author} {\bibfnamefont {Y.}~\bibnamefont
  {Tsaturyan}}, \bibinfo {author} {\bibfnamefont {A.}~\bibnamefont {Barg}},
  \bibinfo {author} {\bibfnamefont {E.~S.}\ \bibnamefont {Polzik}},\ and\
  \bibinfo {author} {\bibfnamefont {A.}~\bibnamefont {Schliesser}},\ }\bibfield
   {title} {\bibinfo {title} {Ultracoherent nanomechanical resonators via soft
  clamping and dissipation dilution},\ }\href
  {https://doi.org/10.1038/nnano.2017.101} {\bibfield  {journal} {\bibinfo
  {journal} {Nature Nanotech}\ }\textbf {\bibinfo {volume} {12}},\ \bibinfo
  {pages} {776} (\bibinfo {year} {2017})}\BibitemShut {NoStop}%
\bibitem [{\citenamefont {Fedorov}\ \emph {et~al.}(2019)\citenamefont
  {Fedorov}, \citenamefont {Engelsen}, \citenamefont {Ghadimi}, \citenamefont
  {Bereyhi}, \citenamefont {Schilling}, \citenamefont {Wilson},\ and\
  \citenamefont {Kippenberg}}]{Fedorov2019}%
  \BibitemOpen
  \bibfield  {author} {\bibinfo {author} {\bibfnamefont {S.~A.}\ \bibnamefont
  {Fedorov}}, \bibinfo {author} {\bibfnamefont {N.~J.}\ \bibnamefont
  {Engelsen}}, \bibinfo {author} {\bibfnamefont {A.~H.}\ \bibnamefont
  {Ghadimi}}, \bibinfo {author} {\bibfnamefont {M.~J.}\ \bibnamefont
  {Bereyhi}}, \bibinfo {author} {\bibfnamefont {R.}~\bibnamefont {Schilling}},
  \bibinfo {author} {\bibfnamefont {D.~J.}\ \bibnamefont {Wilson}},\ and\
  \bibinfo {author} {\bibfnamefont {T.~J.}\ \bibnamefont {Kippenberg}},\
  }\bibfield  {title} {\bibinfo {title} {Generalized dissipation dilution in
  strained mechanical resonators},\ }\href
  {https://doi.org/10.1103/PhysRevB.99.054107} {\bibfield  {journal} {\bibinfo
  {journal} {Phys. Rev. B}\ }\textbf {\bibinfo {volume} {99}},\ \bibinfo
  {pages} {054107} (\bibinfo {year} {2019})}\BibitemShut {NoStop}%
\bibitem [{\citenamefont {Engelsen}\ \emph {et~al.}(2021)\citenamefont
  {Engelsen}, \citenamefont {Agrawal},\ and\ \citenamefont
  {Wilson}}]{Engelsen2021}%
  \BibitemOpen
  \bibfield  {author} {\bibinfo {author} {\bibfnamefont {N.~J.}\ \bibnamefont
  {Engelsen}}, \bibinfo {author} {\bibfnamefont {A.~R.}\ \bibnamefont
  {Agrawal}},\ and\ \bibinfo {author} {\bibfnamefont {D.~J.}\ \bibnamefont
  {Wilson}},\ }\bibfield  {title} {\bibinfo {title} {Ultra-{{High-Q
  Nanomechanics Through Dissipation Dilution}}: {{Trends}} and
  {{Perspectives}}},\ }in\ \href
  {https://doi.org/10.1109/Transducers50396.2021.9495394} {\emph {\bibinfo
  {booktitle} {2021 21st {{Int}}. {{Conf}}. {{Solid-State Sens}}. {{Actuators
  Microsyst}}. {{Transducers}}}}}\ (\bibinfo  {publisher} {{IEEE}},\ \bibinfo
  {address} {{Orlando, FL, USA}},\ \bibinfo {year} {2021})\ pp.\ \bibinfo
  {pages} {201--205}\BibitemShut {NoStop}%
\bibitem [{\citenamefont {Sementilli}\ \emph {et~al.}(2022)\citenamefont
  {Sementilli}, \citenamefont {Romero},\ and\ \citenamefont
  {Bowen}}]{Sementilli2022}%
  \BibitemOpen
  \bibfield  {author} {\bibinfo {author} {\bibfnamefont {L.}~\bibnamefont
  {Sementilli}}, \bibinfo {author} {\bibfnamefont {E.}~\bibnamefont {Romero}},\
  and\ \bibinfo {author} {\bibfnamefont {W.~P.}\ \bibnamefont {Bowen}},\
  }\bibfield  {title} {\bibinfo {title} {Nanomechanical {{Dissipation}} and
  {{Strain Engineering}}},\ }\href {https://doi.org/10.1002/adfm.202105247}
  {\bibfield  {journal} {\bibinfo  {journal} {Adv Funct Materials}\ }\textbf
  {\bibinfo {volume} {32}},\ \bibinfo {pages} {2105247} (\bibinfo {year}
  {2022})}\BibitemShut {NoStop}%
\bibitem [{\citenamefont {Shin}\ \emph {et~al.}(2022)\citenamefont {Shin},
  \citenamefont {Cupertino}, \citenamefont {{de Jong}}, \citenamefont
  {Steeneken}, \citenamefont {Bessa},\ and\ \citenamefont {Norte}}]{Shin2022}%
  \BibitemOpen
  \bibfield  {author} {\bibinfo {author} {\bibfnamefont {D.}~\bibnamefont
  {Shin}}, \bibinfo {author} {\bibfnamefont {A.}~\bibnamefont {Cupertino}},
  \bibinfo {author} {\bibfnamefont {M.~H.~J.}\ \bibnamefont {{de Jong}}},
  \bibinfo {author} {\bibfnamefont {P.~G.}\ \bibnamefont {Steeneken}}, \bibinfo
  {author} {\bibfnamefont {M.~A.}\ \bibnamefont {Bessa}},\ and\ \bibinfo
  {author} {\bibfnamefont {R.~A.}\ \bibnamefont {Norte}},\ }\bibfield  {title}
  {\bibinfo {title} {Spiderweb {{Nanomechanical Resonators}} via {{Bayesian
  Optimization}}: {{Inspired}} by {{Nature}} and {{Guided}} by {{Machine
  Learning}}},\ }\href {https://doi.org/10.1002/adma.202106248} {\bibfield
  {journal} {\bibinfo  {journal} {Advanced Materials}\ }\textbf {\bibinfo
  {volume} {34}},\ \bibinfo {pages} {2106248} (\bibinfo {year}
  {2022})}\BibitemShut {NoStop}%
\bibitem [{\citenamefont {Bereyhi}\ \emph {et~al.}(2022)\citenamefont
  {Bereyhi}, \citenamefont {Beccari}, \citenamefont {Groth}, \citenamefont
  {Fedorov}, \citenamefont {Arabmoheghi}, \citenamefont {Kippenberg},\ and\
  \citenamefont {Engelsen}}]{Bereyhi2022a}%
  \BibitemOpen
  \bibfield  {author} {\bibinfo {author} {\bibfnamefont {M.~J.}\ \bibnamefont
  {Bereyhi}}, \bibinfo {author} {\bibfnamefont {A.}~\bibnamefont {Beccari}},
  \bibinfo {author} {\bibfnamefont {R.}~\bibnamefont {Groth}}, \bibinfo
  {author} {\bibfnamefont {S.~A.}\ \bibnamefont {Fedorov}}, \bibinfo {author}
  {\bibfnamefont {A.}~\bibnamefont {Arabmoheghi}}, \bibinfo {author}
  {\bibfnamefont {T.~J.}\ \bibnamefont {Kippenberg}},\ and\ \bibinfo {author}
  {\bibfnamefont {N.~J.}\ \bibnamefont {Engelsen}},\ }\bibfield  {title}
  {\bibinfo {title} {Hierarchical tensile structures with ultralow mechanical
  dissipation},\ }\href {https://doi.org/10.1038/s41467-022-30586-z} {\bibfield
   {journal} {\bibinfo  {journal} {Nat Commun}\ }\textbf {\bibinfo {volume}
  {13}},\ \bibinfo {pages} {3097} (\bibinfo {year} {2022})},\ \Eprint
  {https://arxiv.org/abs/2103.09785} {arxiv:2103.09785} \BibitemShut {NoStop}%
\bibitem [{\citenamefont {Li}\ \emph {et~al.}(2023)\citenamefont {Li},
  \citenamefont {Xu}, \citenamefont {Norte}, \citenamefont {Arag{\'o}n},
  \citenamefont {{van Keulen}}, \citenamefont {Alijani},\ and\ \citenamefont
  {Steeneken}}]{Li2023}%
  \BibitemOpen
  \bibfield  {author} {\bibinfo {author} {\bibfnamefont {Z.}~\bibnamefont
  {Li}}, \bibinfo {author} {\bibfnamefont {M.}~\bibnamefont {Xu}}, \bibinfo
  {author} {\bibfnamefont {R.~A.}\ \bibnamefont {Norte}}, \bibinfo {author}
  {\bibfnamefont {A.~M.}\ \bibnamefont {Arag{\'o}n}}, \bibinfo {author}
  {\bibfnamefont {F.}~\bibnamefont {{van Keulen}}}, \bibinfo {author}
  {\bibfnamefont {F.}~\bibnamefont {Alijani}},\ and\ \bibinfo {author}
  {\bibfnamefont {P.~G.}\ \bibnamefont {Steeneken}},\ }\bibfield  {title}
  {\bibinfo {title} {Tuning the {{Q-factor}} of nanomechanical string
  resonators by torsion support design},\ }\href
  {https://doi.org/10.1063/5.0133177} {\bibfield  {journal} {\bibinfo
  {journal} {Appl. Phys. Lett.}\ }\textbf {\bibinfo {volume} {122}},\ \bibinfo
  {pages} {013501} (\bibinfo {year} {2023})}\BibitemShut {NoStop}%
\bibitem [{\citenamefont {Zhang}\ \emph {et~al.}(2001)\citenamefont {Zhang},
  \citenamefont {Tanaka}, \citenamefont {Kanki}, \citenamefont {Choi},\ and\
  \citenamefont {Kawai}}]{Zhang2001}%
  \BibitemOpen
  \bibfield  {author} {\bibinfo {author} {\bibfnamefont {J.}~\bibnamefont
  {Zhang}}, \bibinfo {author} {\bibfnamefont {H.}~\bibnamefont {Tanaka}},
  \bibinfo {author} {\bibfnamefont {T.}~\bibnamefont {Kanki}}, \bibinfo
  {author} {\bibfnamefont {J.-H.}\ \bibnamefont {Choi}},\ and\ \bibinfo
  {author} {\bibfnamefont {T.}~\bibnamefont {Kawai}},\ }\bibfield  {title}
  {\bibinfo {title} {Strain effect and the phase diagram of
  {{La1}}-{{xBaxMnO3}} thin films},\ }\href
  {https://doi.org/10.1103/PhysRevB.64.184404} {\bibfield  {journal} {\bibinfo
  {journal} {Phys. Rev. B}\ }\textbf {\bibinfo {volume} {64}},\ \bibinfo
  {pages} {184404} (\bibinfo {year} {2001})}\BibitemShut {NoStop}%
\bibitem [{\citenamefont {{MacManus-Driscoll}}\ \emph
  {et~al.}(2008)\citenamefont {{MacManus-Driscoll}}, \citenamefont {Zerrer},
  \citenamefont {Wang}, \citenamefont {Yang}, \citenamefont {Yoon},
  \citenamefont {Fouchet}, \citenamefont {Yu}, \citenamefont {Blamire},\ and\
  \citenamefont {Jia}}]{MacManus-Driscoll2008}%
  \BibitemOpen
  \bibfield  {author} {\bibinfo {author} {\bibfnamefont {J.~L.}\ \bibnamefont
  {{MacManus-Driscoll}}}, \bibinfo {author} {\bibfnamefont {P.}~\bibnamefont
  {Zerrer}}, \bibinfo {author} {\bibfnamefont {H.}~\bibnamefont {Wang}},
  \bibinfo {author} {\bibfnamefont {H.}~\bibnamefont {Yang}}, \bibinfo {author}
  {\bibfnamefont {J.}~\bibnamefont {Yoon}}, \bibinfo {author} {\bibfnamefont
  {A.}~\bibnamefont {Fouchet}}, \bibinfo {author} {\bibfnamefont
  {R.}~\bibnamefont {Yu}}, \bibinfo {author} {\bibfnamefont {M.~G.}\
  \bibnamefont {Blamire}},\ and\ \bibinfo {author} {\bibfnamefont
  {Q.}~\bibnamefont {Jia}},\ }\bibfield  {title} {\bibinfo {title} {Strain
  control and spontaneous phase ordering in vertical nanocomposite
  heteroepitaxial thin films},\ }\href {https://doi.org/10.1038/nmat2124}
  {\bibfield  {journal} {\bibinfo  {journal} {Nature Mater}\ }\textbf {\bibinfo
  {volume} {7}},\ \bibinfo {pages} {314} (\bibinfo {year} {2008})}\BibitemShut
  {NoStop}%
\bibitem [{\citenamefont {Cao}\ and\ \citenamefont {Wu}(2011)}]{Cao2011}%
  \BibitemOpen
  \bibfield  {author} {\bibinfo {author} {\bibfnamefont {J.}~\bibnamefont
  {Cao}}\ and\ \bibinfo {author} {\bibfnamefont {J.}~\bibnamefont {Wu}},\
  }\bibfield  {title} {\bibinfo {title} {Strain effects in low-dimensional
  transition metal oxides},\ }\href
  {https://doi.org/10.1016/j.mser.2010.08.001} {\bibfield  {journal} {\bibinfo
  {journal} {Materials Science and Engineering: R: Reports}\ }\textbf {\bibinfo
  {volume} {71}},\ \bibinfo {pages} {35} (\bibinfo {year} {2011})}\BibitemShut
  {NoStop}%
\bibitem [{\citenamefont {Mattoni}\ \emph {et~al.}(2018)\citenamefont
  {Mattoni}, \citenamefont {Manca}, \citenamefont {Hadjimichael}, \citenamefont
  {Zubko}, \citenamefont {Van Der~Torren}, \citenamefont {Yin}, \citenamefont
  {Catalano}, \citenamefont {Gibert}, \citenamefont {Maccherozzi},
  \citenamefont {Liu}, \citenamefont {Dhesi},\ and\ \citenamefont
  {Caviglia}}]{Mattoni2018}%
  \BibitemOpen
  \bibfield  {author} {\bibinfo {author} {\bibfnamefont {G.}~\bibnamefont
  {Mattoni}}, \bibinfo {author} {\bibfnamefont {N.}~\bibnamefont {Manca}},
  \bibinfo {author} {\bibfnamefont {M.}~\bibnamefont {Hadjimichael}}, \bibinfo
  {author} {\bibfnamefont {P.}~\bibnamefont {Zubko}}, \bibinfo {author}
  {\bibfnamefont {A.~J.~H.}\ \bibnamefont {Van Der~Torren}}, \bibinfo {author}
  {\bibfnamefont {C.}~\bibnamefont {Yin}}, \bibinfo {author} {\bibfnamefont
  {S.}~\bibnamefont {Catalano}}, \bibinfo {author} {\bibfnamefont
  {M.}~\bibnamefont {Gibert}}, \bibinfo {author} {\bibfnamefont
  {F.}~\bibnamefont {Maccherozzi}}, \bibinfo {author} {\bibfnamefont
  {Y.}~\bibnamefont {Liu}}, \bibinfo {author} {\bibfnamefont {S.~S.}\
  \bibnamefont {Dhesi}},\ and\ \bibinfo {author} {\bibfnamefont {A.~D.}\
  \bibnamefont {Caviglia}},\ }\bibfield  {title} {\bibinfo {title} {Light
  control of the nanoscale phase separation in heteroepitaxial nickelates},\
  }\href {https://doi.org/10.1103/PhysRevMaterials.2.085002} {\bibfield
  {journal} {\bibinfo  {journal} {Phys. Rev. Materials}\ }\textbf {\bibinfo
  {volume} {2}},\ \bibinfo {pages} {085002} (\bibinfo {year}
  {2018})}\BibitemShut {NoStop}%
\bibitem [{\citenamefont {Manca}\ \emph {et~al.}(2022)\citenamefont {Manca},
  \citenamefont {Remaggi}, \citenamefont {Plaza}, \citenamefont {Varbaro},
  \citenamefont {Bernini}, \citenamefont {Pellegrino},\ and\ \citenamefont
  {Marr{\'e}}}]{Manca2022}%
  \BibitemOpen
  \bibfield  {author} {\bibinfo {author} {\bibfnamefont {N.}~\bibnamefont
  {Manca}}, \bibinfo {author} {\bibfnamefont {F.}~\bibnamefont {Remaggi}},
  \bibinfo {author} {\bibfnamefont {A.~E.}\ \bibnamefont {Plaza}}, \bibinfo
  {author} {\bibfnamefont {L.}~\bibnamefont {Varbaro}}, \bibinfo {author}
  {\bibfnamefont {C.}~\bibnamefont {Bernini}}, \bibinfo {author} {\bibfnamefont
  {L.}~\bibnamefont {Pellegrino}},\ and\ \bibinfo {author} {\bibfnamefont
  {D.}~\bibnamefont {Marr{\'e}}},\ }\bibfield  {title} {\bibinfo {title}
  {Stress {{Analysis}} and {{Q}}-{{Factor}} of {{Free}}-{{Standing}}
  ({{La}},{{Sr}}){{MnO}}{\textsubscript{3}} {{Oxide Resonators}}},\ }\href
  {https://doi.org/10.1002/smll.202202768} {\bibfield  {journal} {\bibinfo
  {journal} {Small}\ }\textbf {\bibinfo {volume} {18}},\ \bibinfo {pages}
  {2202768} (\bibinfo {year} {2022})}\BibitemShut {NoStop}%
\bibitem [{\citenamefont {Sambri}\ \emph {et~al.}(2020)\citenamefont {Sambri},
  \citenamefont {Scuderi}, \citenamefont {Guarino}, \citenamefont {Gennaro},
  \citenamefont {Erlandsen}, \citenamefont {Dahm}, \citenamefont {Bj{\o}rlig},
  \citenamefont {Christensen}, \citenamefont {Capua}, \citenamefont {Ventura},
  \citenamefont {Uccio}, \citenamefont {Mirabella}, \citenamefont {Nicotra},
  \citenamefont {Spinella}, \citenamefont {Jespersen},\ and\ \citenamefont
  {Granozio}}]{Sambri2020}%
  \BibitemOpen
  \bibfield  {author} {\bibinfo {author} {\bibfnamefont {A.}~\bibnamefont
  {Sambri}}, \bibinfo {author} {\bibfnamefont {M.}~\bibnamefont {Scuderi}},
  \bibinfo {author} {\bibfnamefont {A.}~\bibnamefont {Guarino}}, \bibinfo
  {author} {\bibfnamefont {E.~D.}\ \bibnamefont {Gennaro}}, \bibinfo {author}
  {\bibfnamefont {R.}~\bibnamefont {Erlandsen}}, \bibinfo {author}
  {\bibfnamefont {R.~T.}\ \bibnamefont {Dahm}}, \bibinfo {author}
  {\bibfnamefont {A.~V.}\ \bibnamefont {Bj{\o}rlig}}, \bibinfo {author}
  {\bibfnamefont {D.~V.}\ \bibnamefont {Christensen}}, \bibinfo {author}
  {\bibfnamefont {R.~D.}\ \bibnamefont {Capua}}, \bibinfo {author}
  {\bibfnamefont {B.~D.}\ \bibnamefont {Ventura}}, \bibinfo {author}
  {\bibfnamefont {U.~S.~D.}\ \bibnamefont {Uccio}}, \bibinfo {author}
  {\bibfnamefont {S.}~\bibnamefont {Mirabella}}, \bibinfo {author}
  {\bibfnamefont {G.}~\bibnamefont {Nicotra}}, \bibinfo {author} {\bibfnamefont
  {C.}~\bibnamefont {Spinella}}, \bibinfo {author} {\bibfnamefont {T.~S.}\
  \bibnamefont {Jespersen}},\ and\ \bibinfo {author} {\bibfnamefont {F.~M.}\
  \bibnamefont {Granozio}},\ }\bibfield  {title} {\bibinfo {title}
  {Self-{{Formed}}, {{Conducting
  LaAlO}}{\textsubscript{3}}/{{SrTiO}}{\textsubscript{3}}
  {{Micro}}-{{Membranes}}},\ }\href {https://doi.org/10.1002/adfm.201909964}
  {\bibfield  {journal} {\bibinfo  {journal} {Adv. Funct. Mater.}\ }\textbf
  {\bibinfo {volume} {30}},\ \bibinfo {pages} {1909964} (\bibinfo {year}
  {2020})}\BibitemShut {NoStop}%
\bibitem [{\citenamefont {Manca}\ \emph {et~al.}(2019)\citenamefont {Manca},
  \citenamefont {Mattoni}, \citenamefont {Pelassa}, \citenamefont {Venstra},
  \citenamefont {{van der Zant}},\ and\ \citenamefont {Caviglia}}]{Manca2019a}%
  \BibitemOpen
  \bibfield  {author} {\bibinfo {author} {\bibfnamefont {N.}~\bibnamefont
  {Manca}}, \bibinfo {author} {\bibfnamefont {G.}~\bibnamefont {Mattoni}},
  \bibinfo {author} {\bibfnamefont {M.}~\bibnamefont {Pelassa}}, \bibinfo
  {author} {\bibfnamefont {W.~J.}\ \bibnamefont {Venstra}}, \bibinfo {author}
  {\bibfnamefont {H.~S.~J.}\ \bibnamefont {{van der Zant}}},\ and\ \bibinfo
  {author} {\bibfnamefont {A.~D.}\ \bibnamefont {Caviglia}},\ }\bibfield
  {title} {\bibinfo {title} {Large {{Tunability}} of {{Strain}} in
  {{WO}}{\textsubscript{3}} {{Single-Crystal Microresonators Controlled}} by
  {{Exposure}} to {{H}}{\textsubscript{2}} {{Gas}}},\ }\href
  {https://doi.org/10.1021/acsami.9b14501} {\bibfield  {journal} {\bibinfo
  {journal} {ACS Appl. Mater. Interfaces}\ }\textbf {\bibinfo {volume} {11}},\
  \bibinfo {pages} {44438} (\bibinfo {year} {2019})}\BibitemShut {NoStop}%
\bibitem [{\citenamefont {Peacock}\ and\ \citenamefont
  {Robertson}(2002)}]{Peacock2002}%
  \BibitemOpen
  \bibfield  {author} {\bibinfo {author} {\bibfnamefont {P.~W.}\ \bibnamefont
  {Peacock}}\ and\ \bibinfo {author} {\bibfnamefont {J.}~\bibnamefont
  {Robertson}},\ }\bibfield  {title} {\bibinfo {title} {Band offsets and
  {{Schottky}} barrier heights of high dielectric constant oxides},\ }\href
  {https://doi.org/10.1063/1.1506388} {\bibfield  {journal} {\bibinfo
  {journal} {J. Appl. Phys.}\ }\textbf {\bibinfo {volume} {92}},\ \bibinfo
  {pages} {4712} (\bibinfo {year} {2002})}\BibitemShut {NoStop}%
\bibitem [{\citenamefont {Choi}\ \emph {et~al.}(2012)\citenamefont {Choi},
  \citenamefont {Rouleau}, \citenamefont {Seo}, \citenamefont {Luo},
  \citenamefont {Zhou}, \citenamefont {Fister}, \citenamefont {Eastman},
  \citenamefont {Fuoss}, \citenamefont {Fong}, \citenamefont {Tischler},
  \citenamefont {Eres}, \citenamefont {Chisholm},\ and\ \citenamefont
  {Lee}}]{Choi2012}%
  \BibitemOpen
  \bibfield  {author} {\bibinfo {author} {\bibfnamefont {W.~S.}\ \bibnamefont
  {Choi}}, \bibinfo {author} {\bibfnamefont {C.~M.}\ \bibnamefont {Rouleau}},
  \bibinfo {author} {\bibfnamefont {S.~S.~A.}\ \bibnamefont {Seo}}, \bibinfo
  {author} {\bibfnamefont {Z.}~\bibnamefont {Luo}}, \bibinfo {author}
  {\bibfnamefont {H.}~\bibnamefont {Zhou}}, \bibinfo {author} {\bibfnamefont
  {T.~T.}\ \bibnamefont {Fister}}, \bibinfo {author} {\bibfnamefont {J.~A.}\
  \bibnamefont {Eastman}}, \bibinfo {author} {\bibfnamefont {P.~H.}\
  \bibnamefont {Fuoss}}, \bibinfo {author} {\bibfnamefont {D.~D.}\ \bibnamefont
  {Fong}}, \bibinfo {author} {\bibfnamefont {J.~Z.}\ \bibnamefont {Tischler}},
  \bibinfo {author} {\bibfnamefont {G.}~\bibnamefont {Eres}}, \bibinfo {author}
  {\bibfnamefont {M.~F.}\ \bibnamefont {Chisholm}},\ and\ \bibinfo {author}
  {\bibfnamefont {H.~N.}\ \bibnamefont {Lee}},\ }\bibfield  {title} {\bibinfo
  {title} {Atomic {{Layer Engineering}} of {{Perovskite Oxides}} for
  {{Chemically Sharp Heterointerfaces}}},\ }\href
  {https://doi.org/10.1002/adma.201202691} {\bibfield  {journal} {\bibinfo
  {journal} {Advanced Materials}\ }\textbf {\bibinfo {volume} {24}},\ \bibinfo
  {pages} {6423} (\bibinfo {year} {2012})}\BibitemShut {NoStop}%
\bibitem [{\citenamefont {Plaza}\ \emph {et~al.}(2021)\citenamefont {Plaza},
  \citenamefont {Manca}, \citenamefont {Bernini}, \citenamefont {Marr{\'e}},\
  and\ \citenamefont {Pellegrino}}]{Plaza2021}%
  \BibitemOpen
  \bibfield  {author} {\bibinfo {author} {\bibfnamefont {A.~E.}\ \bibnamefont
  {Plaza}}, \bibinfo {author} {\bibfnamefont {N.}~\bibnamefont {Manca}},
  \bibinfo {author} {\bibfnamefont {C.}~\bibnamefont {Bernini}}, \bibinfo
  {author} {\bibfnamefont {D.}~\bibnamefont {Marr{\'e}}},\ and\ \bibinfo
  {author} {\bibfnamefont {L.}~\bibnamefont {Pellegrino}},\ }\bibfield  {title}
  {\bibinfo {title} {The role of etching anisotropy in the fabrication of
  freestanding oxide microstructures on {{SrTiO}}{\textsubscript{3}}(100),
  {{SrTiO}}{\textsubscript{3}}(110), and {{SrTiO}}{\textsubscript{3}}(111)
  substrates},\ }\href {https://doi.org/10.1063/5.0056524} {\bibfield
  {journal} {\bibinfo  {journal} {Appl. Phys. Lett.}\ }\textbf {\bibinfo
  {volume} {119}},\ \bibinfo {pages} {033504} (\bibinfo {year}
  {2021})}\BibitemShut {NoStop}%
\bibitem [{\citenamefont {Schmid}\ \emph {et~al.}(2016)\citenamefont {Schmid},
  \citenamefont {Villanueva},\ and\ \citenamefont {Roukes}}]{Schmid2016}%
  \BibitemOpen
  \bibfield  {author} {\bibinfo {author} {\bibfnamefont {S.}~\bibnamefont
  {Schmid}}, \bibinfo {author} {\bibfnamefont {L.~G.}\ \bibnamefont
  {Villanueva}},\ and\ \bibinfo {author} {\bibfnamefont {M.~L.}\ \bibnamefont
  {Roukes}},\ }\href {https://doi.org/10.1007/978-3-319-28691-4} {\emph
  {\bibinfo {title} {Fundamentals of {{Nanomechanical Resonators}}}}}\
  (\bibinfo  {publisher} {{Springer International Publishing}},\ \bibinfo
  {address} {{Cham}},\ \bibinfo {year} {2016})\BibitemShut {NoStop}%
\bibitem [{\citenamefont {Harrison}\ \emph {et~al.}(2004)\citenamefont
  {Harrison}, \citenamefont {Redfern}, \citenamefont {Buckley},\ and\
  \citenamefont {Salje}}]{Harrison2004}%
  \BibitemOpen
  \bibfield  {author} {\bibinfo {author} {\bibfnamefont {R.~J.}\ \bibnamefont
  {Harrison}}, \bibinfo {author} {\bibfnamefont {S.~A.~T.}\ \bibnamefont
  {Redfern}}, \bibinfo {author} {\bibfnamefont {A.}~\bibnamefont {Buckley}},\
  and\ \bibinfo {author} {\bibfnamefont {E.~K.~H.}\ \bibnamefont {Salje}},\
  }\bibfield  {title} {\bibinfo {title} {Application of real-time, stroboscopic
  x-ray diffraction with dynamical mechanical analysis to characterize the
  motion of ferroelastic domain walls},\ }\href
  {https://doi.org/10.1063/1.1639949} {\bibfield  {journal} {\bibinfo
  {journal} {J. Appl. Phys.}\ }\textbf {\bibinfo {volume} {95}},\ \bibinfo
  {pages} {1706} (\bibinfo {year} {2004})}\BibitemShut {NoStop}%
\bibitem [{\citenamefont {Carpenter}\ \emph {et~al.}(2010)\citenamefont
  {Carpenter}, \citenamefont {Buckley}, \citenamefont {Taylor},\ and\
  \citenamefont {Darling}}]{Carpenter2010b}%
  \BibitemOpen
  \bibfield  {author} {\bibinfo {author} {\bibfnamefont {M.~A.}\ \bibnamefont
  {Carpenter}}, \bibinfo {author} {\bibfnamefont {A.}~\bibnamefont {Buckley}},
  \bibinfo {author} {\bibfnamefont {P.~A.}\ \bibnamefont {Taylor}},\ and\
  \bibinfo {author} {\bibfnamefont {T.~W.}\ \bibnamefont {Darling}},\
  }\bibfield  {title} {\bibinfo {title} {Elastic relaxations associated with
  the {{Pm3m-R3c}} transition in {{LaAlO}}{\textsubscript{3}}: {{III}}.
  {{Superattenuation}} of acoustic resonances},\ }\href
  {https://doi.org/10.1088/0953-8984/22/3/035405} {\bibfield  {journal}
  {\bibinfo  {journal} {J. Phys.: Condens. Matter}\ }\textbf {\bibinfo {volume}
  {22}},\ \bibinfo {pages} {035405} (\bibinfo {year} {2010})}\BibitemShut
  {NoStop}%
\bibitem [{\citenamefont {Harbola}\ \emph {et~al.}(2021)\citenamefont
  {Harbola}, \citenamefont {Crossley}, \citenamefont {Hong}, \citenamefont
  {Lu}, \citenamefont {Birkh{\"o}lzer}, \citenamefont {Hikita},\ and\
  \citenamefont {Hwang}}]{Harbola2021}%
  \BibitemOpen
  \bibfield  {author} {\bibinfo {author} {\bibfnamefont {V.}~\bibnamefont
  {Harbola}}, \bibinfo {author} {\bibfnamefont {S.}~\bibnamefont {Crossley}},
  \bibinfo {author} {\bibfnamefont {S.~S.}\ \bibnamefont {Hong}}, \bibinfo
  {author} {\bibfnamefont {D.}~\bibnamefont {Lu}}, \bibinfo {author}
  {\bibfnamefont {Y.~A.}\ \bibnamefont {Birkh{\"o}lzer}}, \bibinfo {author}
  {\bibfnamefont {Y.}~\bibnamefont {Hikita}},\ and\ \bibinfo {author}
  {\bibfnamefont {H.~Y.}\ \bibnamefont {Hwang}},\ }\bibfield  {title} {\bibinfo
  {title} {Strain {{Gradient Elasticity}} in {{SrTiO}}{\textsubscript{3}}
  {{Membranes}}: {{Bending}} versus {{Stretching}}},\ }\href
  {https://doi.org/10.1021/acs.nanolett.0c04787} {\bibfield  {journal}
  {\bibinfo  {journal} {Nano Lett.}\ }\textbf {\bibinfo {volume} {21}},\
  \bibinfo {pages} {2470} (\bibinfo {year} {2021})}\BibitemShut {NoStop}%
\bibitem [{\citenamefont {Villanueva}\ and\ \citenamefont
  {Schmid}(2014)}]{Villanueva2014}%
  \BibitemOpen
  \bibfield  {author} {\bibinfo {author} {\bibfnamefont {L.~G.}\ \bibnamefont
  {Villanueva}}\ and\ \bibinfo {author} {\bibfnamefont {S.}~\bibnamefont
  {Schmid}},\ }\bibfield  {title} {\bibinfo {title} {Evidence of {{Surface
  Loss}} as {{Ubiquitous Limiting Damping Mechanism}} in {{SiN Micro-}} and
  {{Nanomechanical Resonators}}},\ }\href
  {https://doi.org/10.1103/PhysRevLett.113.227201} {\bibfield  {journal}
  {\bibinfo  {journal} {Phys. Rev. Lett.}\ }\textbf {\bibinfo {volume} {113}},\
  \bibinfo {pages} {227201} (\bibinfo {year} {2014})}\BibitemShut {NoStop}%
\bibitem [{\citenamefont {Mazini}\ \emph {et~al.}(2022)\citenamefont {Mazini},
  \citenamefont {Favre}, \citenamefont {Ariosa},\ and\ \citenamefont
  {Faccio}}]{Mazini2022}%
  \BibitemOpen
  \bibfield  {author} {\bibinfo {author} {\bibfnamefont {M.}~\bibnamefont
  {Mazini}}, \bibinfo {author} {\bibfnamefont {S.}~\bibnamefont {Favre}},
  \bibinfo {author} {\bibfnamefont {D.}~\bibnamefont {Ariosa}},\ and\ \bibinfo
  {author} {\bibfnamefont {R.}~\bibnamefont {Faccio}},\ }\bibfield  {title}
  {\bibinfo {title} {Substrate and thickness influence on
  {{YBa2Cu3O}}\{7-{$\Delta\rbrace$} Thin Films Grown by \vphantom\}{{PLD}}
  deposition},\ }\href {https://doi.org/10.1007/s00339-022-06202-8} {\bibfield
  {journal} {\bibinfo  {journal} {Appl. Phys. A}\ }\textbf {\bibinfo {volume}
  {128}},\ \bibinfo {pages} {1111} (\bibinfo {year} {2022})}\BibitemShut
  {NoStop}%
\bibitem [{\citenamefont {Benzi}\ \emph {et~al.}(2004)\citenamefont {Benzi},
  \citenamefont {Bottizzo},\ and\ \citenamefont {Rizzi}}]{Benzi2004}%
  \BibitemOpen
  \bibfield  {author} {\bibinfo {author} {\bibfnamefont {P.}~\bibnamefont
  {Benzi}}, \bibinfo {author} {\bibfnamefont {E.}~\bibnamefont {Bottizzo}},\
  and\ \bibinfo {author} {\bibfnamefont {N.}~\bibnamefont {Rizzi}},\ }\bibfield
   {title} {\bibinfo {title} {Oxygen determination from cell dimensions in
  {{YBCO}} superconductors},\ }\href
  {https://doi.org/10.1016/j.jcrysgro.2004.05.082} {\bibfield  {journal}
  {\bibinfo  {journal} {Journal of Crystal Growth}\ }\textbf {\bibinfo {volume}
  {269}},\ \bibinfo {pages} {625} (\bibinfo {year} {2004})}\BibitemShut
  {NoStop}%
\bibitem [{\citenamefont {Chakrabarty}\ \emph {et~al.}(2019)\citenamefont
  {Chakrabarty}, \citenamefont {Lindeman}, \citenamefont {Bumble},
  \citenamefont {Kleinsasser}, \citenamefont {Holmes},\ and\ \citenamefont
  {Cunnane}}]{Chakrabarty2019}%
  \BibitemOpen
  \bibfield  {author} {\bibinfo {author} {\bibfnamefont {A.}~\bibnamefont
  {Chakrabarty}}, \bibinfo {author} {\bibfnamefont {M.~A.}\ \bibnamefont
  {Lindeman}}, \bibinfo {author} {\bibfnamefont {B.}~\bibnamefont {Bumble}},
  \bibinfo {author} {\bibfnamefont {A.~W.}\ \bibnamefont {Kleinsasser}},
  \bibinfo {author} {\bibfnamefont {W.~A.}\ \bibnamefont {Holmes}},\ and\
  \bibinfo {author} {\bibfnamefont {D.}~\bibnamefont {Cunnane}},\ }\bibfield
  {title} {\bibinfo {title} {Operation of {{YBCO}} kinetic-inductance
  bolometers for outer solar system missions},\ }\href
  {https://doi.org/10.1063/1.5089143} {\bibfield  {journal} {\bibinfo
  {journal} {Appl. Phys. Lett.}\ }\textbf {\bibinfo {volume} {114}},\ \bibinfo
  {pages} {132601} (\bibinfo {year} {2019})}\BibitemShut {NoStop}%
\bibitem [{\citenamefont {Maspero}\ \emph {et~al.}(2021)\citenamefont
  {Maspero}, \citenamefont {Gatani}, \citenamefont {Cuccurullo},\ and\
  \citenamefont {Bertacco}}]{Maspero2021}%
  \BibitemOpen
  \bibfield  {author} {\bibinfo {author} {\bibfnamefont {F.}~\bibnamefont
  {Maspero}}, \bibinfo {author} {\bibfnamefont {G.}~\bibnamefont {Gatani}},
  \bibinfo {author} {\bibfnamefont {S.}~\bibnamefont {Cuccurullo}},\ and\
  \bibinfo {author} {\bibfnamefont {R.}~\bibnamefont {Bertacco}},\ }\bibfield
  {title} {\bibinfo {title} {{{MEMS Magnetometer Using Magnetic Flux
  Concentrators}} and {{Permanent Magnets}}},\ }in\ \href
  {https://doi.org/10.1109/MEMS51782.2021.9375441} {\emph {\bibinfo {booktitle}
  {2021 {{IEEE}} 34th {{Int}}. {{Conf}}. {{Micro Electro Mech}}. {{Syst}}.
  {{MEMS}}}}}\ (\bibinfo  {publisher} {{IEEE}},\ \bibinfo {address}
  {{Gainesville, FL, USA}},\ \bibinfo {year} {2021})\ pp.\ \bibinfo {pages}
  {374--377}\BibitemShut {NoStop}%
\end{thebibliography}%




\end{document}